\documentclass[letterpaper, preprint, paper,11pt]{AAS}

\usepackage{mathtools}
\usepackage{gensymb}
\usepackage{float}
\usepackage{graphicx}
\usepackage{xcolor}
\usepackage{tabularx}
\usepackage{adjustbox}

\usepackage{bm}
\usepackage{amsmath}
\usepackage{subfigure}
\usepackage{comment}
\usepackage{overcite}
\usepackage{footnpag}			      	
\usepackage{amssymb}
\usepackage{subfigure}   
\usepackage{multicol}
\usepackage{multirow}
\usepackage{booktabs}
\usepackage{pstool}             
\usepackage[colorlinks]{hyperref}

\usepackage{amsthm}

\PaperNumber{25-623}

\makeatletter
\newcommand{\ostar}{\mathbin{\mathpalette\make@circled*}}
\newcommand{\make@circled}[2]{%
  \ooalign{$\m@th#1\smallbigcirc{#1}$\cr\hidewidth$\m@th#1#2$\hidewidth\cr}%
}
\newcommand{\smallbigcirc}[1]{%
  \vcenter{\hbox{\scalebox{0.77778}{$\m@th#1\bigcirc$}}}%
}
\makeatother
\usepackage{subfigure}

\begin{document}

\title{Periodic orbit tracking in cislunar space: A finite-horizon approach}

\author{ Mohammed Atallah\thanks{PhD Student, Department of Aerospace Engineering, Iowa State University, IA 50011, USA. email: matallah@iastate.edu} 
\ and Simone Servadio\thanks{Assistant Professor, Department of Aerospace Engineering, Iowa State University, IA 50011, USA. email: servadio@iastate.edu}
}

 \maketitle
\begin{abstract}
This paper presents a Nonlinear Model Predictive Control (NMPC) scheme for maintaining a spacecraft within a specified family of periodic orbits near the libration points in cislunar space. Unlike traditional approaches that track a predefined reference orbit, the proposed method designs an optimal trajectory that keeps the spacecraft within the orbit family, regardless of the initial reference. The Circular Restricted Three-Body Problem (CR3BP) is used to model the system dynamics. First, the Pseudo-Arclength Continuation (PAC) method is employed to compute the members of each orbit family. Then, the state of each member is parameterized by two variables: one defining the orbit and the other specifying the location along it. These computed states are then fit to a Multivariate Polynomial Regression (MPR) model. An NMPC framework is developed to generate the optimal reference trajectory and compute the corresponding velocity impulses for trajectory tracking. The control system is integrated with a Extended Kalman Filter (EKF) observer that estimates the spacecraft’s relative state. Numerical simulations are conducted for Lyapunov, halo, and near-rectilinear halo orbits near \(\mathcal{L}_1\) and \(\mathcal{L}_2\). The results demonstrate a significant reduction in fuel consumption compared to conventional tracking methods.
\end{abstract}
 
\section{Introduction}

The growing interest in deep-space exploration has brought increased focus to cislunar space, particularly due to its strategic role in enabling low-energy trajectory design~[\citen{koon2000dynamical}]. However, this region is characterized by chaotic dynamics resulting from its complex multi-body gravitational environment. These conditions highlight the importance of maintaining bounded motion—especially periodic and quasi-periodic orbits near libration points~[\citen{wilmer2024preliminary}]. Over the past few decades, cislunar orbital dynamics have been extensively studied, with recent missions placing greater emphasis on periodic orbits near libration points as key assets. One notable example is the ARTEMIS mission, which demonstrated successful station-keeping near both the \(\mathcal{L}_1\) and \(\mathcal{L}_2\) points~[\citen{woodard2009artemis}]. Another is the Lunar Orbiter Platform-Gateway (LOP-G), an international collaborative effort aimed at assembling a space station in lunar orbit~[\citen{merri2018lunar}]. Furthermore, cislunar operations are integral to NASA’s long-term roadmap for Mars exploration~[\citen{national2016nasa}]. The increasing significance of this region is underscored by the more than thirty missions planned over the next decade~[\citen{baker2024comprehensive}], with periodic orbits identified as prime locations for mission staging and sustained operations~[\citen{whitley2016options}].

The Circular Restricted Three-Body Problem (CR3BP) remains the most widely used model for designing cislunar trajectories, offering tractable approximations for transfer and station-keeping strategies~[\citen{richardson1980analytic}]. It has been applied in various mission contexts: Pritchett et al. developed low-energy transfers between periodic orbits~[\citen{pritchett2018impulsive}]; Singh et al. leveraged \(\mathcal{L}_1\) Halo orbits for designing low-thrust lunar transfers~[\citen{singh2020low}]; and Davis et al. combined CR3BP-based periodic orbits with n-body dynamics for station-keeping analysis~[\citen{davis2017stationkeeping}]. Van et al. used the Pseudo-Arclength Continuation (PAC) method to compute complete families of periodic orbits~[\citen{van2016tadpole}]. The CR3BP has also proven effective for constellation maintenance~[\citen{wilmer2021cislunar}] and comparative studies of rescue operations across different orbit types~[\citen{fay2024investigation}]. While the Bicircular Restricted Four-Body Problem (BCR4BP) offers improved accuracy—as demonstrated in~[\citen{negri2020generalizing, oshima2022multiple, wilmer2021lagrangian}]—its increased computational cost and lack of closed-form solutions present challenges for practical mission design.

Several studies have applied Model Predictive Control (MPC) to cislunar mission scenarios. In~[\citen{salzo2023design}], an MPC scheme is proposed for satellite formation flying around Halo orbits using the CR3BP model. In~[\citen{sanchez2020chance}], a robust guidance and control system employs chance-constrained MPC for cislunar rendezvous missions, combining the CR3BP with a disturbance estimator to account for perturbations. In~[\citen{capannolo2023model}], a Model Predictive Guidance and Control approach is introduced for formation reconfiguration in NRHOs, incorporating collision avoidance and limited-thrust constraints. A different approach is presented in~[\citen{quartullo2023periodic}], where a periodic MPC scheme is used to track Halo orbits for station-keeping and rendezvous missions. Unlike the commonly used CR3BP, this work adopts the Elliptic Restricted Three-Body Problem (ER3BP) to model satellite dynamics. Similarly,~[\citen{berning2020suboptimal}] applies the ER3BP in a Nonlinear MPC (NMPC) framework for tracking NRHOs, aiming to stabilize the spacecraft around a predefined reference orbit while accommodating thrust limitations. Servadio et al. made use of the Koopman Operator [\citen{servadio2021koopman}] to develop a computationally fast NMPC in cislunar [\citen{servadio2022dynamics}].

In this study, we develop an NMPC scheme to maintain a spacecraft within a specified family of periodic orbits near libration points in cislunar space. Unlike conventional methods that rely on tracking a predefined reference orbit, the proposed approach formulates an optimal trajectory ensuring the spacecraft remains within the periodic orbit family, regardless of the initial reference. To achieve this, the reference state in the scheme is represented by a Multivariate Polynomial Regression (MPR) model. The CR3BP is employed to model the motion dynamics in cislunar. The Pseudo-Arclength Continuation (PAC) method is first used to compute the members of each orbit family; these states are then parameterized using two variables: one to identify the orbit and another to determine the location within it. The computed members, along with their parameterized states, are fitted to an MPR model. The NMPC framework utilizes this model to generate the optimal reference trajectory and calculate the necessary velocity impulses for trajectory tracking. Numerical simulations for Lyapunov, halo, and near-rectilinear halo orbits around \(\mathcal{L}_1\) and \(\mathcal{L}_2\) demonstrate significant reductions in fuel consumption compared to traditional tracking methods.

The rest of the paper is organized as follows: Section~2 presents the mathematical model of the CR3BP. Section~3 shows the evaluation of the periodic orbits near $\mathcal{L}_1$ and $\mathcal{L}_2$. Section~4 presents the development of the MPR models. Section~5 develops the NMPC scheme. Section~7 presents and discusses the results of the numerical simulations and demonstrates the applicability of the proposed methodology. To further validate the applicability of the proposed scheme, the system is integrated with Extended Kalman Filter (EKF). Section~8 concludes the paper.

\section{Mathematical Model of the CR3BP}
{The CR3BP model is used to represent the translational motion of the spacecraft in cislunar space, assuming that the Earth and Moon are modeled as point masses, the Moon's motion relative to the Earth is circular, and the only forces acting on the spacecraft are the gravitational attractions of the Earth and the Moon.}

In the conventional form of the CR3BP model, the time and states are dimensionless. The rotating coordinates centered at the barycenter of the Earth-Moon system represent the reference frame, where \(i_x\) is the unit vector in the direction from Earth and the Moon, \(i_y\) is the unit vector in the direction of the in-plane motion of the Moon, and \(i_z\) is the unit vector that completes the set according to the right-hand rule. The mathematical model of the CR3BP is presented as follows:
\begin{equation}
\begin{aligned}
& \ddot{x}=2 \dot{y}+x-\dfrac{(1-\mu)(x+\mu)}{r_1^3}-\dfrac{\mu[x-(1-\mu)]}{r_2^3} \\
& \ddot{y}=-2 \dot{x}+y-\dfrac{(1-\mu) y}{r_1^3}-\dfrac{\mu y}{r_2^3} \\
& \ddot{z}=-\dfrac{(1-\mu) z}{r_1^3}-\dfrac{\mu z}{r_2^3}
\end{aligned}
\label{eq:CR3BP}  
\end{equation}
where, \(x\), \(y\), and \(z\) denote the components of the dimensionless position vector in \(i_x\), \(i_y\), and \(i_z\) of the spacecraft, receptively. \(\mu = 0.01215\) denotes the dimensionless mass of the Moon. \(r_1\) and \(r_2\) are the relative distances measured from the spacecraft to Earth, and to the Moon, respectively.

\section{Periodic Orbits Near $\mathcal{L}_1$ and $\mathcal{L}_2$}

There are five equilibrium points in the CR3BP mode, at which the gravitational forces exerted by the Earth and the Moon are balanced. These points are known as libration points. This study concerns of the first two points, \(\mathcal{L}_1\) and \(\mathcal{L}_2\). There are three well-known periodic orbit families near these points: {Lyapunov Orbits (LOs)}, which exist in two-dimensional space~[\citen{henon1969numerical}], {Halo Orbits (HOs)} and {Near-Rectilinear Halo Orbits (NRHOs)}, which exist in three-dimensional space~[\citen{breakwell1979halo}]. 

{Due to the nonlinearity and chaotic behavior of the CR3BP, periodic orbits are computed through a series of iterative steps. First, the coordinates are transformed to the libration point, and the equations are linearized to derive an analytical expression for the approximate initial state of an arbitrary periodic orbit in the family. Then, a high-order differential correction scheme is applied to refine the initial state and determine the orbit’s period [\citen{servadio2022dynamics}]. Finally, the PAC method is employed to generate other family members using the computed orbit as a seed.} The detailed implementation of these steps is presented in~[\citen{atallah2024advances}].

\subsection{LO Families}
\begin{figure}[H]
    \centering
    \subfigure[Orbits near L1]{%
        \includegraphics[width=0.45\textwidth]{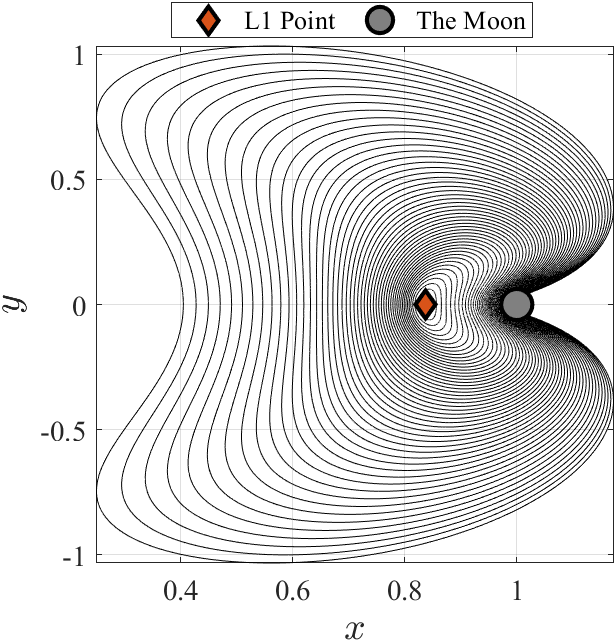}
    }
    \hfill
    \subfigure[Orbits near L2]{%
        \includegraphics[width=0.45\textwidth]{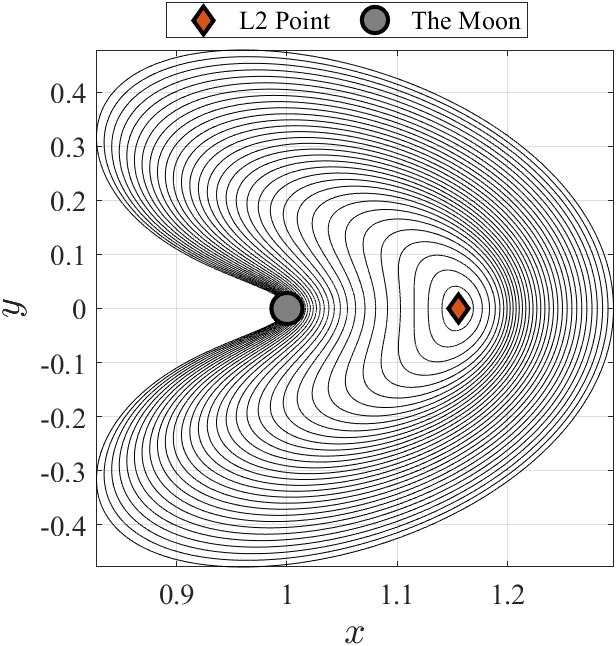}
    }
    \caption{The members of the LO Families.}
    \label{fig:Lyap}
\end{figure}

Figure~\ref{fig:Lyap} shows several LOs near \(\mathcal{L}_1\) and \(\mathcal{L}_2\). A key advantage of this method is its ability to obtain orbits that are close to the Moon.

\subsection{HO and NRHO Families} 
\begin{figure}[H]
    \centering
    \subfigure[Orbits near L1]{%
        \includegraphics[width=0.45\textwidth]{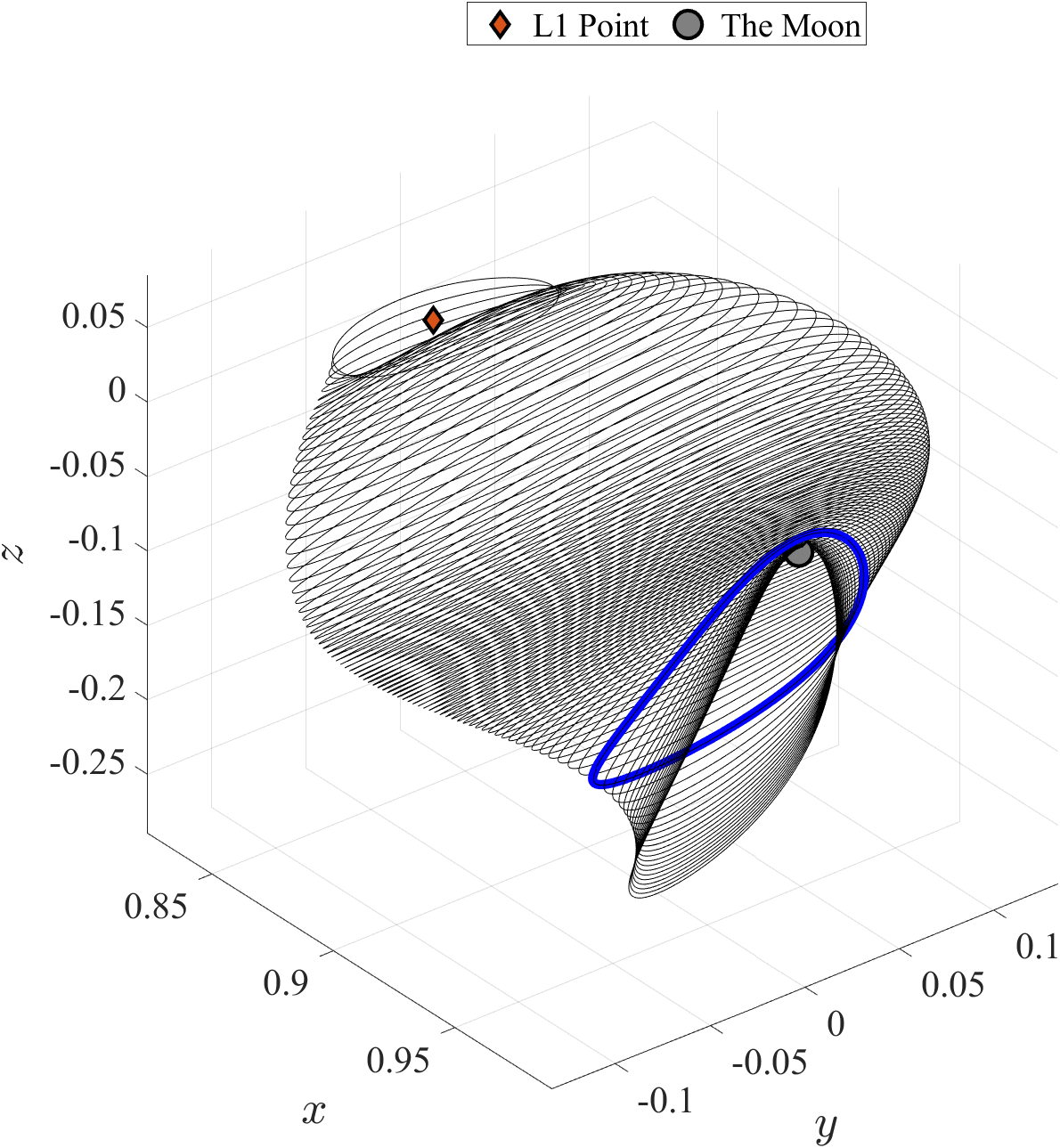}
        \label{fig:Halo_L1}
    }
    \hfill
    \subfigure[Orbits near L2]{%
        \includegraphics[width=0.45\textwidth]{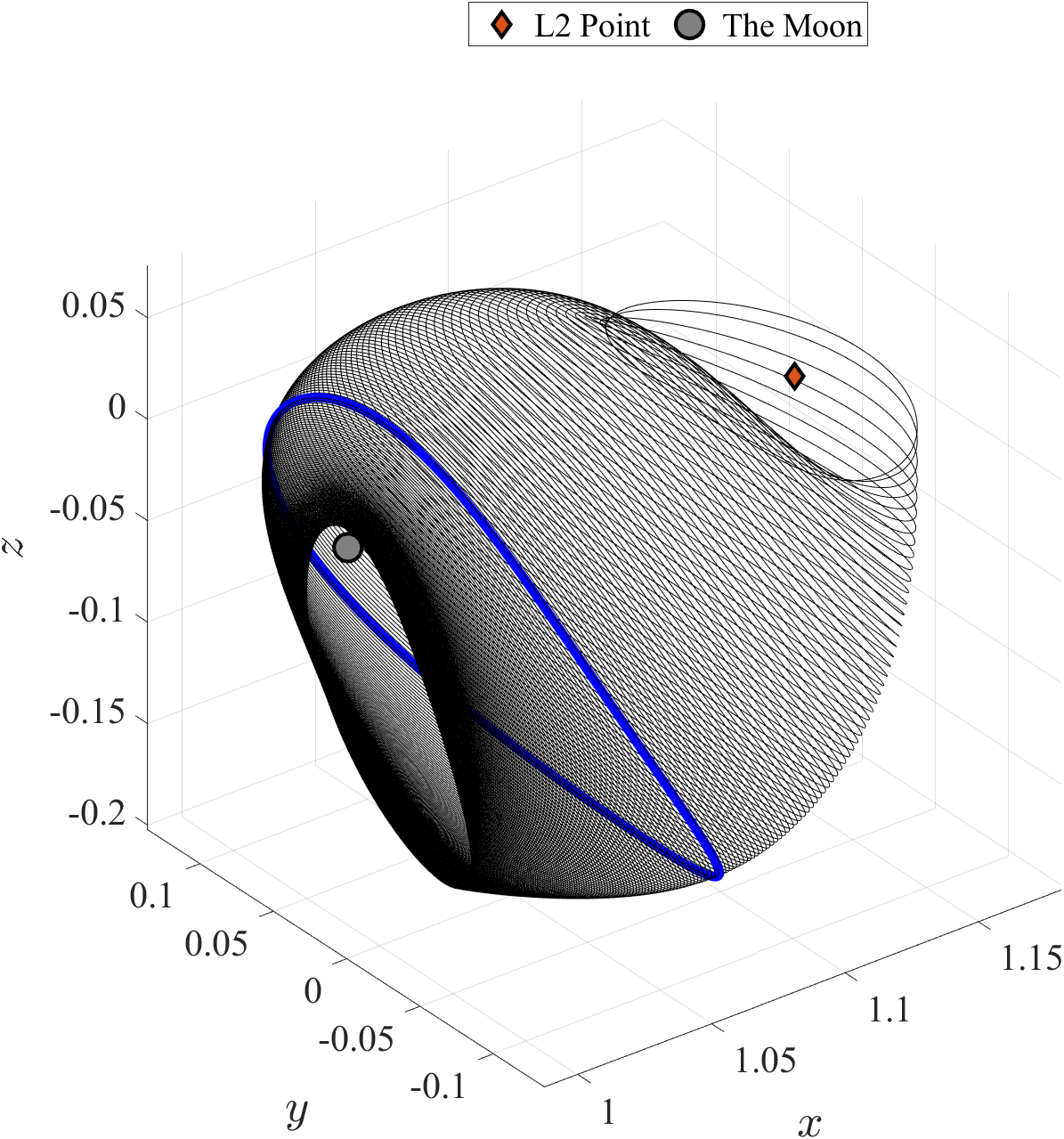}
        \label{fig:Halo_L2}
    }
    \caption{The members of the HO and NRHO families.}
    \label{fig:Halo}
\end{figure}

Similarly, Figure ~\ref{fig:Halo}  shows numerous HOs and NRHOs near \(\mathcal{L}_1\) and \(\mathcal{L}_2\). Specifically, Figure~\ref{fig:Halo_L1} displays the \(\mathcal{L}_1\) families, separated by the bold blue orbit, while Figure~\ref{fig:Halo_L2} shows the \(\mathcal{L}_2\) families.

\section{Parametric Regression Models}
\begin{figure}[H]
    \centering
\includegraphics[width=0.3\linewidth]{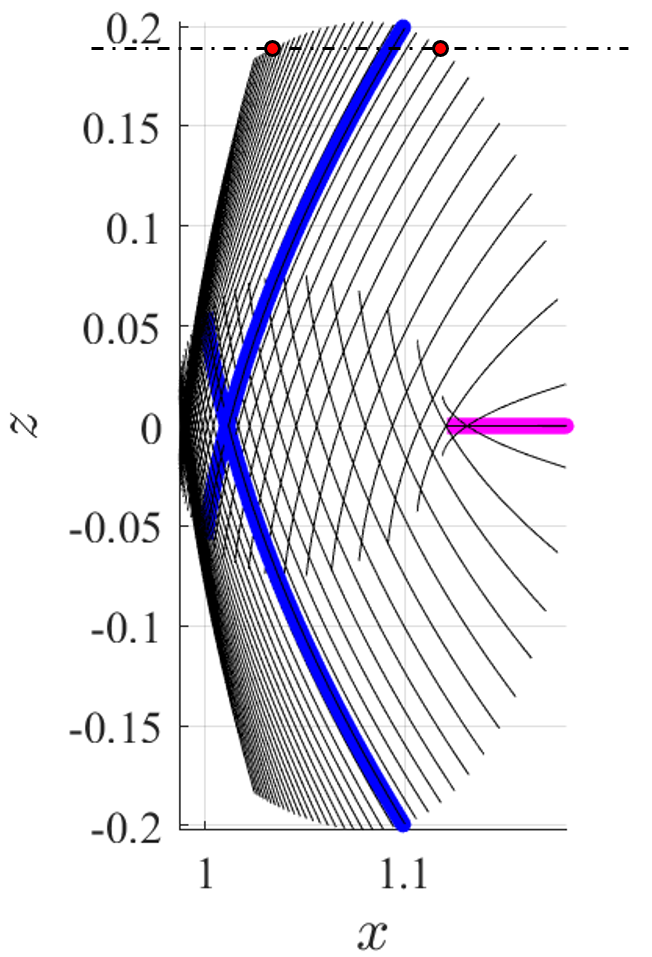}
    \caption{Non-uniqueness of the 
z-components in HOs.}   \label{fig:HOs_Z}
\end{figure}

Each periodic orbit family near the libration points forms a 2-D invariant manifold, which can be represented by the minimal number of independent parameters. {This parameterization reduces the number of states required to represent the motion from four to two, thereby decreasing the computational burden on the control system, particularly in the finite-horizon approach.} {The two independent parameters must be chosen such that each point on the manifold has a unique representation. Therefore, the $z$-component of the HOs cannot be used as an independent parameter, as two orbits may share the same $z$-value, as illustrated in Figure~\ref{fig:HOs_Z}. In addition, the parameters $p_1$ and $p_2$ must be defined such that the state-transition function $f(p_1, p_2)$ is continuous over the region $\mathcal{R}$, where $\mathcal{R} = [L_1, U_1] \times [L_2, U_2] = \{(p_1, p_2) \mid L_1 \leq p_1 \leq U_1,\; L_2 \leq p_2 \leq U_2\}$. Consequently, the Cartesian coordinates $x$ and $y$ cannot be selected as independent parameters.}

\begin{figure}[H]
    \centering
    \subfigure[Orbits near L1]{%
        \includegraphics[width=0.45\textwidth]{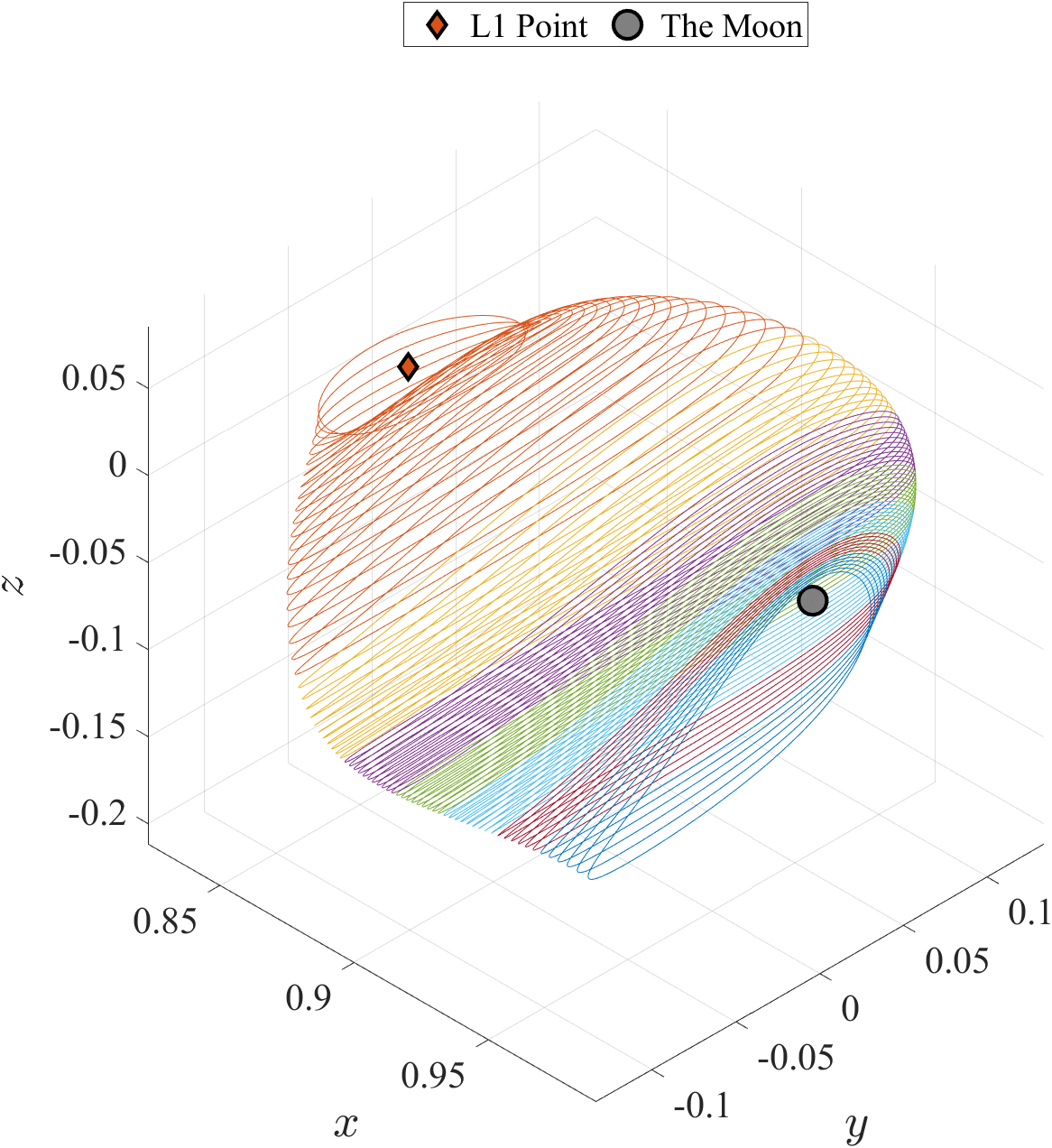}
        \label{fig:Halo_L1_colored}
    }
    \hfill
    \subfigure[Orbits near L2]{%
        \includegraphics[width=0.45\textwidth]{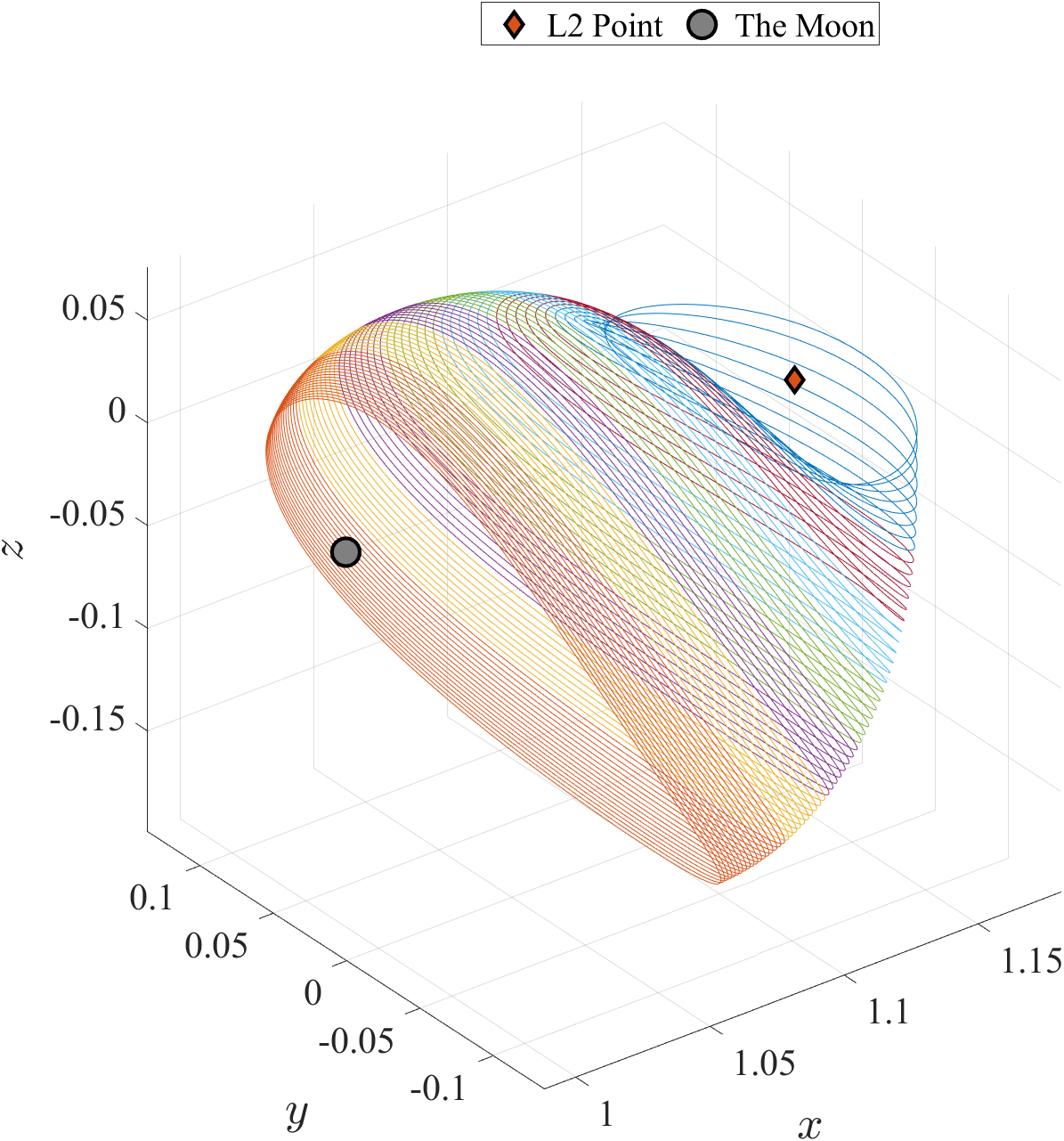}
        \label{fig:Halo_L2_colored}
    }
    \caption{The members of the HO families divided into sub-manifolds.}
    \label{fig:Halo_submanifold}
\end{figure}

One way to parameterize the periodic manifold is to choose one parameter, \(\chi\), to identify the orbit and another parameter, \(\nu\), to specify the location within the orbit. The definition of these two parameters might vary depending on the manifold. However, the parameterization procedure is similar. First, the manifold of each family is divided into multiple sub-manifolds to improve the accuracy of the regression model, as shown in Figure~\ref{fig:Halo_submanifold}, where members within the same sub-manifold are shown in the same color.
Then, a large number of members is computed for each family and used as training data for the model. After that, a multivariate polynomial regression model for each sub-manifold is fit according to the following equation: 
\begin{equation}
{}^{j}\mathbf{X}_i={}^{j}\mathcal{P}_i\left(\chi, \nu\right) = {}^{j}\alpha_{0 i} + \sum_{n_{\chi} + n_{c\nu} + n_{s\nu} \leq N}{}^{j}\alpha_{\left[n_{\chi}, n_{c\nu}, n_{s\nu}\right] i} \; \chi^{n_{\chi}} \cos\left(\nu\right)^{n_{c\nu}} \sin\left(\nu\right)^{n_{s\nu}}
\end{equation}
where \(\mathbf{X}\) denotes the state vector, \(\mathcal{P}\) denotes the PRM, \((\cdot)_i\) denotes the i\textsuperscript{th} element of the vector, \({}^{j}(\cdot)\) denotes the j\textsuperscript{th} sub-manifold, \(N\) denotes the highest order of the polynomial, \(\chi\) is the variable that determines the orbit within the family, \(\nu\) is the variable that determines the location of the spacecraft within the orbit, and \(\alpha_{(\cdot)}\) are the polynomial coefficients. Both \(\chi\) and \(\nu\) are defined based on the family. In the following subsections, we present the parametrization for halo, near-rectilinear, and Lyapunov orbits near \(\mathcal{L}_1\) and \(\mathcal{L}_2\).

\subsection{Parametrization of Lyapunov Orbits}
\begin{figure}[H]
    \centering
    \subfigure[Orbits near \(\mathcal{L}_1\)]{%
        \includegraphics[width=0.45\textwidth]{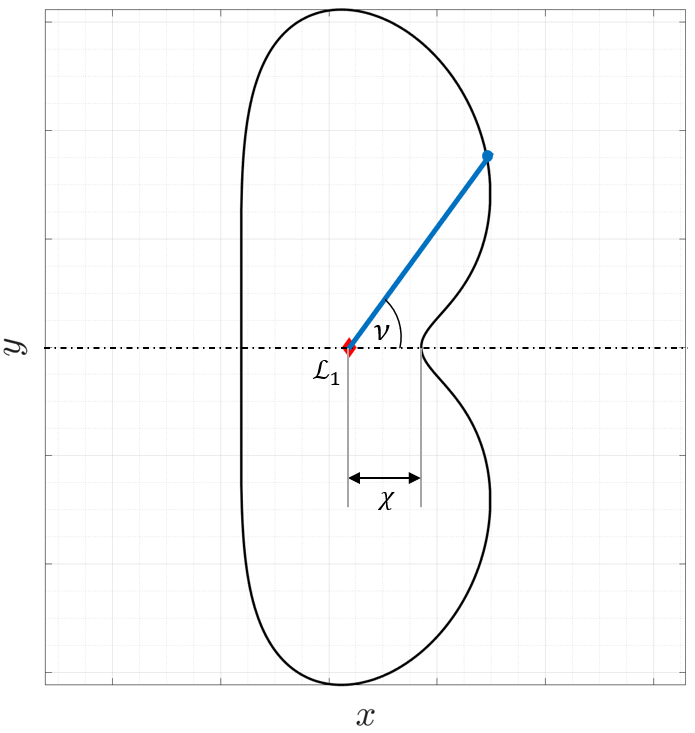}
    }
    \hfill
    \subfigure[Orbits near \(\mathcal{L}_2\)]{%
        \includegraphics[width=0.45\textwidth]{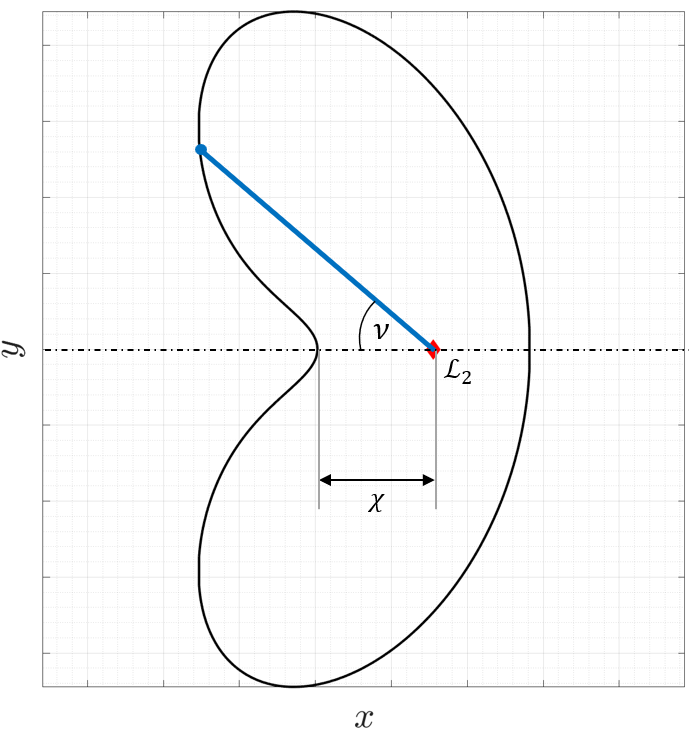}
    }
    \caption{\(\chi\) and \(\nu\) for Lyapunov orbits.}
    \label{fig:chi_nu_Lyap}
\end{figure}

The parameters \(\chi\) and \(\nu\) are defined for Lyapunov orbits such that \(\nu\) represents the angle between the spacecraft's position vector and the \(X\)-axis in the direction that points to the Moon, while \(\chi\) denotes the distance from the libration point to the point on the orbit where \(\nu = 0\), as illustrated in Figure~\ref{fig:chi_nu_Lyap}.

\subsection{Parametrization of Halo orbits and NRHOs}
\begin{figure}[H]
    \centering
        \includegraphics[width=0.95\textwidth]{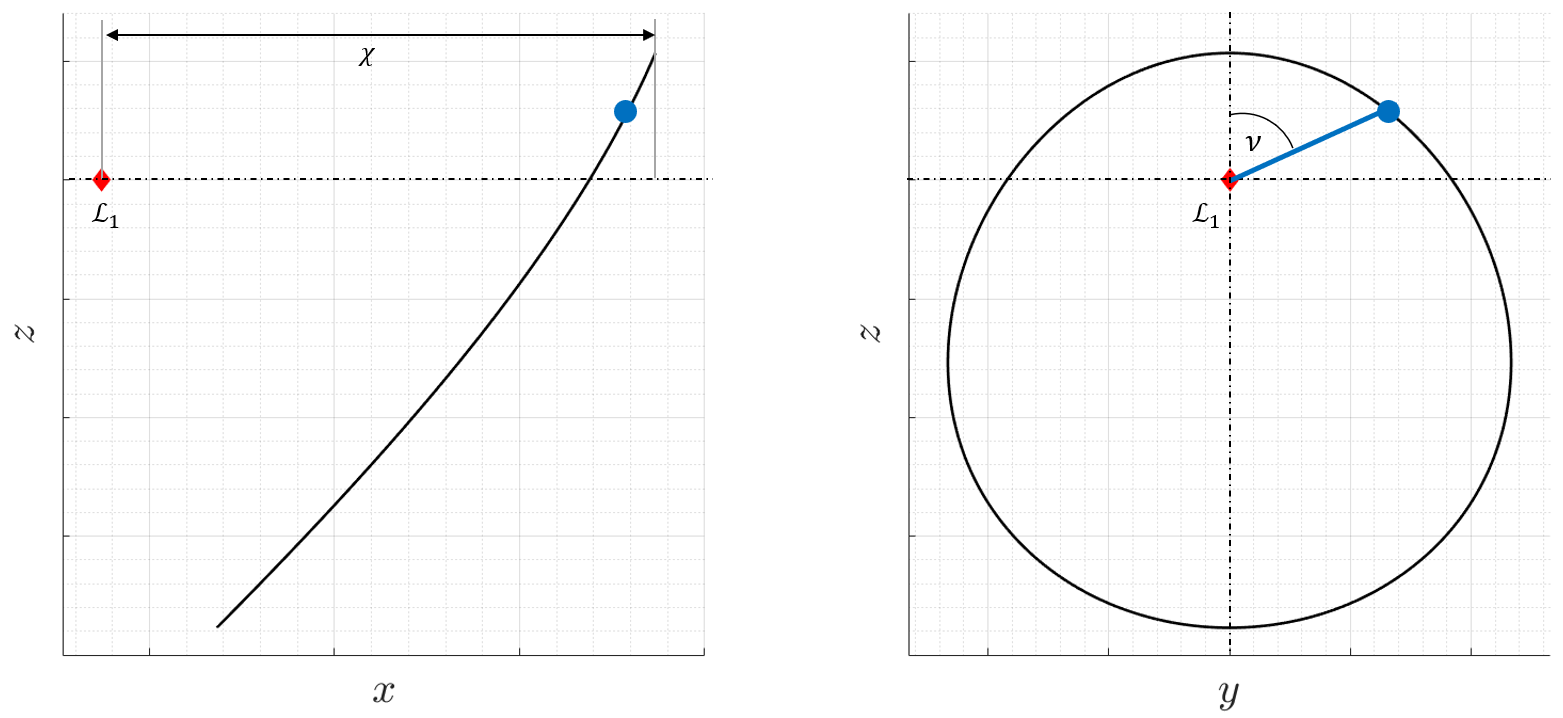}
    \caption{\(\chi\) and \(\nu\) for Halo orbits and NRHOs near \(\mathcal{L}_1\).}
    \label{fig:chi_nu_Halo_L1}
\end{figure}

\begin{figure}[H]
    \centering
        \includegraphics[width=0.95\textwidth]{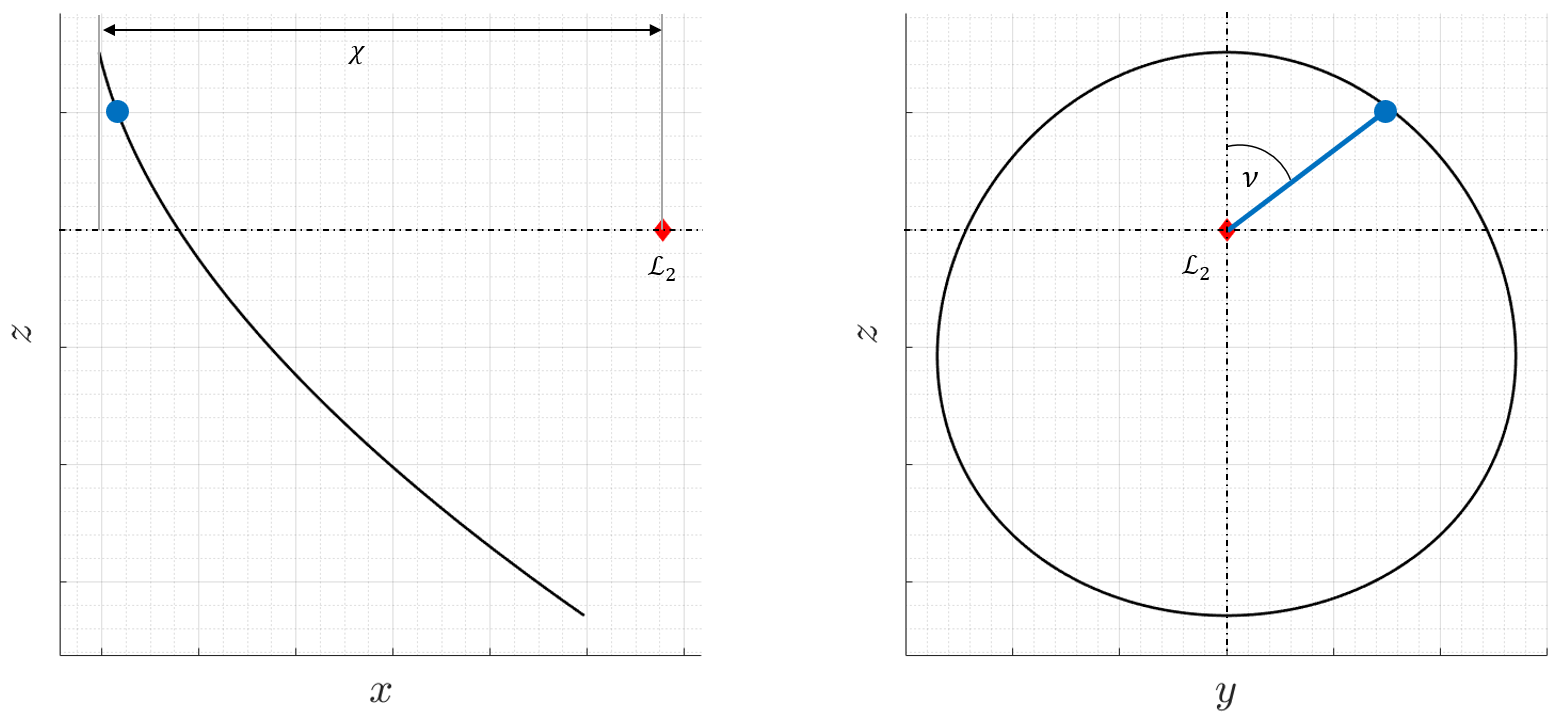}
    \caption{\(\chi\) and \(\nu\) for Halo orbits and NRHOs near \(\mathcal{L}_2\).}
    \label{fig:chi_nu_Halo_L2}
\end{figure}

The parameters \(\chi\) and \(\nu\) are defined for Halo orbits and NRHOs such that \(\chi\) represents the distance between the highest point on the orbit and the corresponding libration point, while \(\nu\) denotes the angle between the spacecraft's projected position in the \(YZ\)-plane and the \(Z\)-axis, as illustrated in Figures~\ref{fig:chi_nu_Halo_L1}~-~\ref{fig:chi_nu_Halo_L2}.

\section{NONLINEAR MODEL PREDICTIVE CONTROL}
The concept of NMPC is to solve a nonlinear optimization problem over a given prediction and control horizon to obtain the optimal control action that guarantees the asymptotic stability of the system. In this section, an NMPC strategy is proposed for trajectory tracking. The objective is to minimize the cost function \(J\), defined as follows:

\begin{equation}
\begin{split}
    J_k =&\;
\sum_{i=1}^{N_c}\Bigg\{\left\|f\left(X_{k+i-1}, \Delta V_{k+i}\right)-\mathcal{P}\left(\chi_k, v_{k+i}\right)\right\|_Q^2 + \left\|\Delta V_{k+i}\right\|_R^2 \\
& + \sum_{j=N_c+1}^{N_p}\left\|f\left(X_{k+j-1}, \Delta V_{k+N_c}\right)-\mathcal{P}\left(\chi_k, v_{k+j}\right)\right\|_Q^2 \Bigg\} \\
& + \left\|f\left(X_{k+N_p+1}, \Delta V_{k+N_c}\right)-\mathcal{P}\left(\chi_k, v_{k+N_p+1}\right)\right\|_{Q_t}^2
\end{split}
\label{eq:J}
\end{equation}
where, \(k\) represents the current time step; \(N_p\) and \(N_c\) are the prediction and control horizons, respectively; \(Q\) and \(Q_t\) are positive semi-definite weighting matrices; and \(R\) is a positive-definite weighting matrix. {$f\left(X_{k+i-1}, \Delta V_{k+i}\right)$ denotes the propagated state using the dynamic model, and $\mathcal{P}\left(\chi_k, v_{k+i}\right)$ represents the reference state computed using the regression model. The objective of the first term is to minimize the error between the propagated and reference states over the prediction horizon, while the second term aims to minimize the control effort over the control horizon. The final term represents the terminal cost function, which ensures the stability of the system’s performance.}

The corresponding constrained optimization problem is formulated as follows:
\begin{equation}
\begin{split}
    & \min _{\Delta V_{k+i}, \chi_k, v_{k+i}} J_k\\
    &\begin{split}
        \text{s.t.} \quad & X_{k} = X_0 \\
        & -\pi \leq \nu_{k+i} < \pi \\
        & \chi_{\min} \leq \chi_k \leq \chi_{\max} \\
        & \Delta V_{\min} \leq \Delta V_{k+i} \leq \Delta V_{\max}
    \end{split}
\end{split}
\label{eq:OP}
\end{equation}
{The objective is to minimize the cost function $J$ subject to input and state constraints. The first constraint on $\nu$ eliminates the discontinuity at $\nu = -\pi$. The second constraint on $\chi$ ensures that the satellite remains within the same sub-manifold it started in. The third constraint on $\Delta V$ ensures that the velocity impulse does not exceed the actuator's hardware limits.}

In Equation~\eqref{eq:J}, \(f\left(X_{k+i-1}, \Delta V_{k+i}\right)\) represents the discrete-time state transition function of the CR3BP, computed using the fourth-order Runge-Kutta method (RK4). To achieve high-precision state propagation, two sampling times are employed: the main sampling time \(T_s\) for \(f(\cdots)\), and a finer propagation sampling time \(\widehat{T}_s\), where \(\widehat{T}_s \leq T_s\). The discretization is performed as follows:
\begin{enumerate}
    \item The state \(X_{k+i-1}\) is first propagated using \(\widehat{T}_s\):
    \begin{equation}
        X^{\prime}_{k+i} = f_{\widehat{T}_s}\left(X_{k+i-1}, \Delta V_{k+i}\right)
    \end{equation}
    \item The state \(X^{\prime}_{k+i}\) is then propagated \(N_T - 1\) additional times with no control input, where \(N_T = T_s / \widehat{T}_s\):
    \begin{equation}
        \begin{split}
            & X^{\prime \prime}_{k+i} =  f_{\widehat{T}_s}\left(X^{\prime}_{k+i}, 0 \right) \\
            & X^{\prime \prime\prime}_{k+i} =  f_{\widehat{T}_s}\left(X^{\prime\prime}_{k+i}, 0 \right) \\
            &\vdots \\
            & X_{k+i} = X^{(N_T)}_{k+i} = f_{\widehat{T}_s}\left(X^{(N_T-1)}_{k+i}, 0 \right)
        \end{split}
    \end{equation}
\end{enumerate}
Although this method is computationally more expensive, it provides more accurate propagation of \(X_{k+i-1}\) for a given velocity impulse \(\Delta V_{k+i}\).

\begin{figure}[H]
    \centering
    \includegraphics[width=0.85\linewidth]{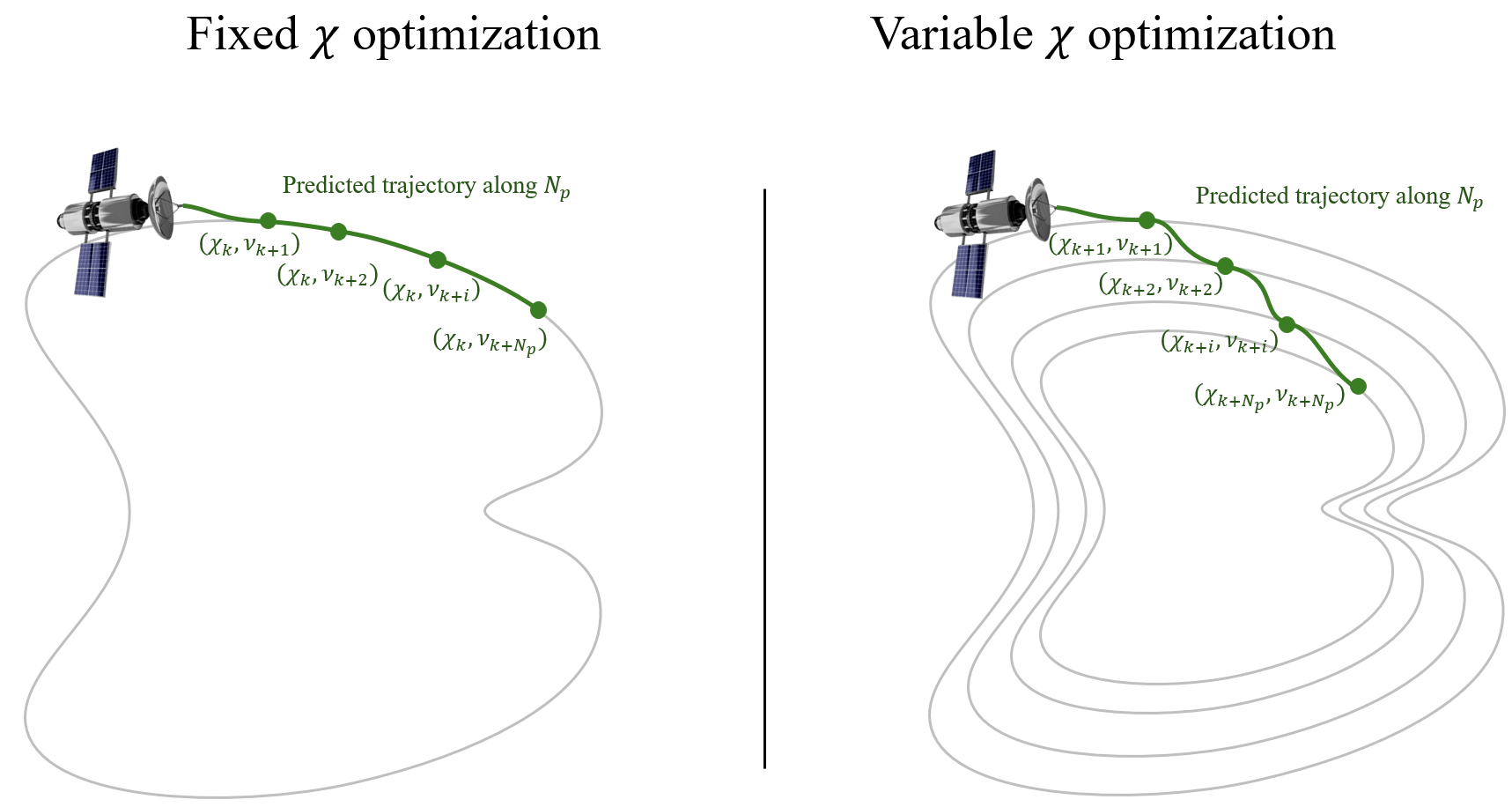}
    \caption{{The Fixed \(\chi\) approach and the variable \(\chi\) approach.}}
    \label{fig:varChi_vs_fxChi}
\end{figure}

It is worth noting that the optimization involves only a single parameter, \(\chi_k\), that determines the optimal target orbit along the prediction horizon. {This means that, at each time step, the optimizer selects a single reference orbit that minimizes the cost function over the prediction horizon.} An alternative approach would be to define a separate parameter \(\chi_{k+i}\) for each step. {In this case, the optimizer searches for a different reference orbit along the prediction horizon. Figure~\ref{fig:varChi_vs_fxChi} depicts the difference between the two approaches.} While the variable-$\chi$ approach may offer higher accuracy due to precise orbit selection at each prediction step, it comes at the cost of increased computational complexity and reduced real-time applicability, it significantly increases computational complexity with only marginal improvements in performance, as demonstrated in the results section.

The optimization problem in Equation~\eqref{eq:OP} is highly non-convex and cannot be solved using traditional quadratic programming methods. Therefore, the CasADi package~[\citen{andersson2019casadi}] is employed to solve for the optimal control sequence. This package has been widely used in NMPC applications in the aerospace field, including those presented in~[\citen{elhesasy2023non, atallah2024analytic, atallah2023development}]. In this study, the NMPC scheme is implemented within a MATLAB environment. {Unlike traditional NMPC schemes, which track a single reference orbit, the approach in this study aims to track an entire family of periodic orbits. In this method, the reference state is treated as an implicit optimization variable along the prediction horizon. As a result, the proposed approach demonstrates improved performance over conventional methods.}

\subsection{Analysis of Prediction and Control Horizon}
The selection of the prediction and control horizons, \(N_p\) and \(N_c\), depends primarily on two factors: the optimality of the computed \(\Delta V\) and the computational complexity. These factors are not necessarily proportional or inversely proportional to \(N_p\) and \(N_c\). Moreover, there is no clear analytical relationship that defines this dependence. Therefore, a common approach is to perform numerical simulations for various values of \(N_p\) and \(N_c\), and evaluate the resulting performance metrics for each case.

\begin{figure}[H]
    \centering
    \includegraphics[width=0.85\linewidth]{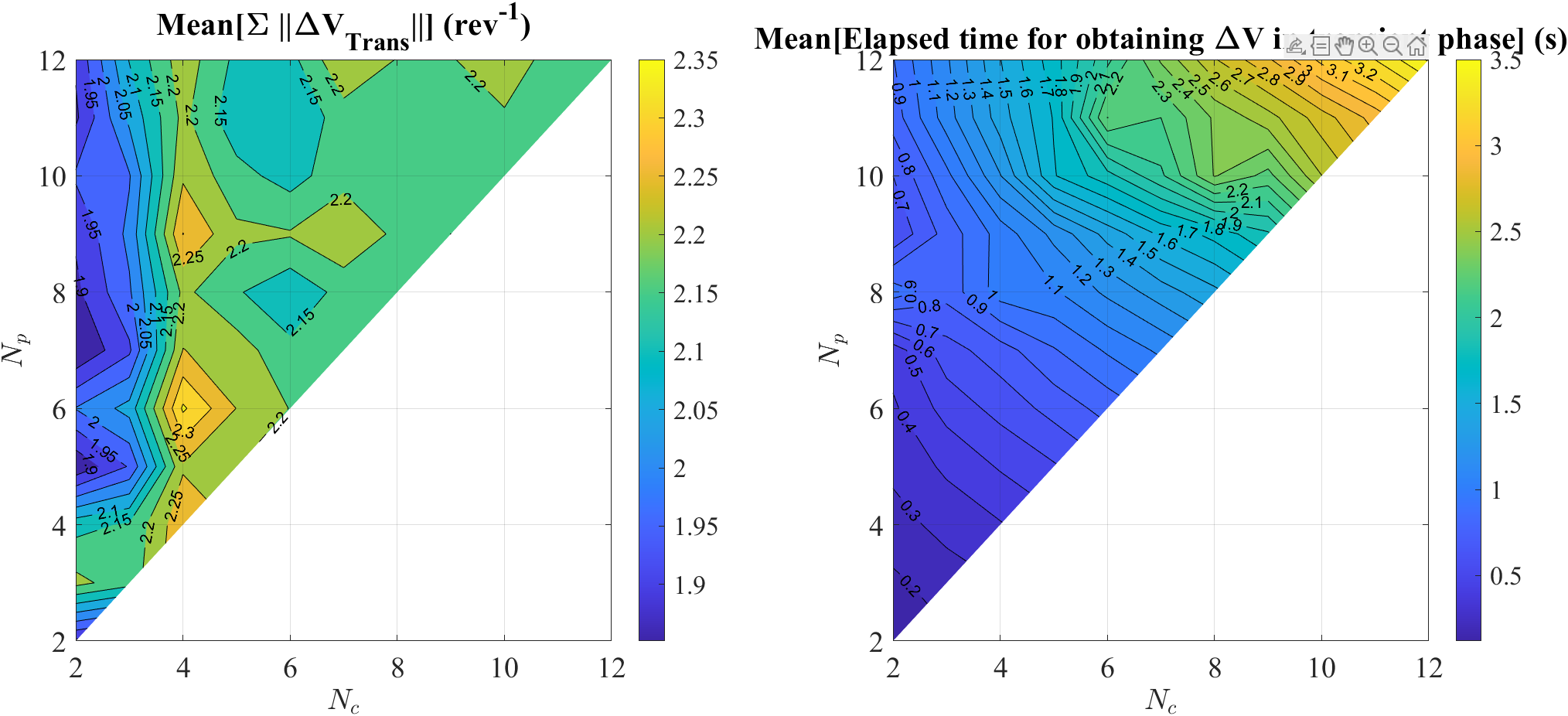}
    \caption{The control efforts and computation time for different \(N_p\) and \(N_c\) in case of LOs.}
    \label{fig:LO_contours}
\end{figure}

\begin{figure}[H]
    \centering
    \includegraphics[width=0.85\linewidth]{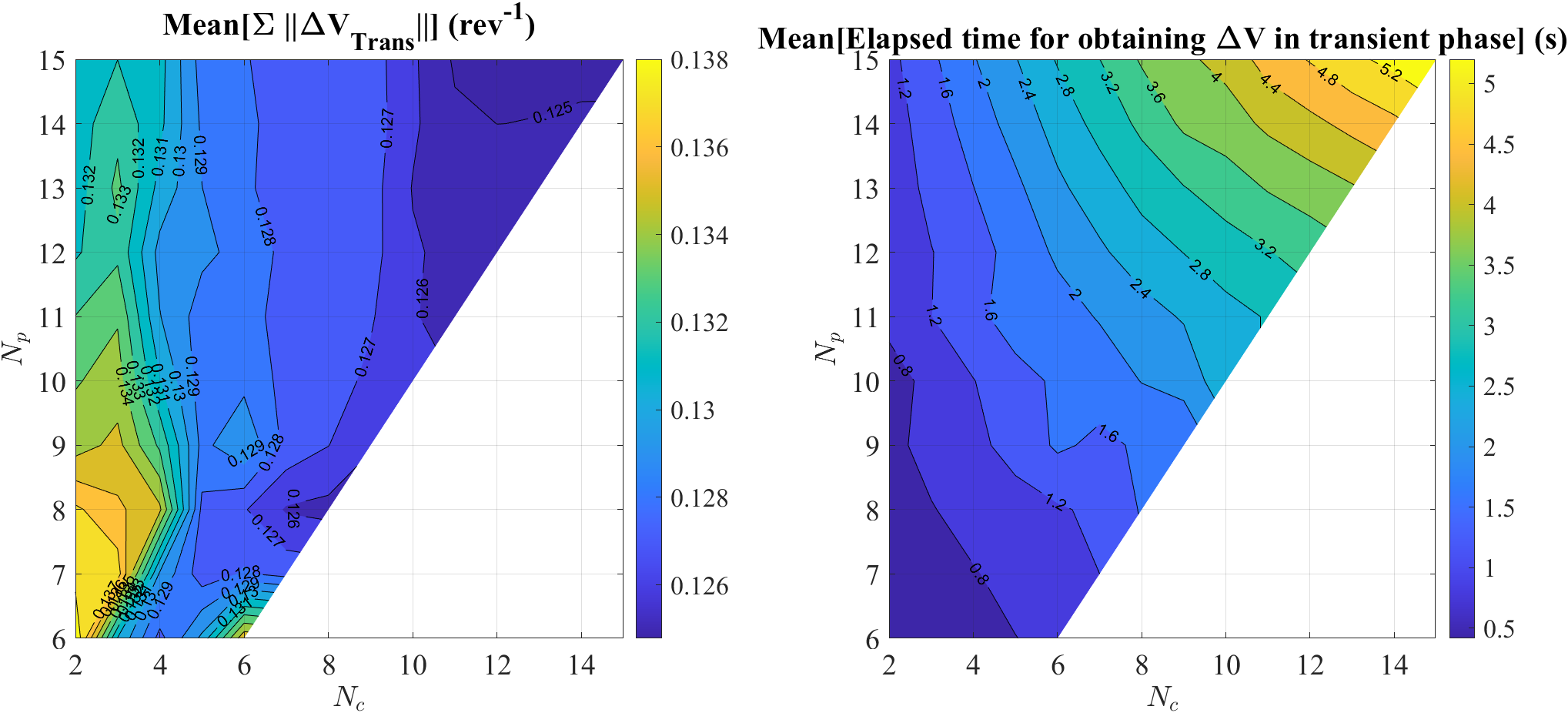}
    \caption{The control efforts and computation time for different \(N_p\) and \(N_c\) in case of HOs.}
    \label{fig:HO_contours}
\end{figure}

\begin{figure}[H]
    \centering
    \includegraphics[width=0.85\linewidth]{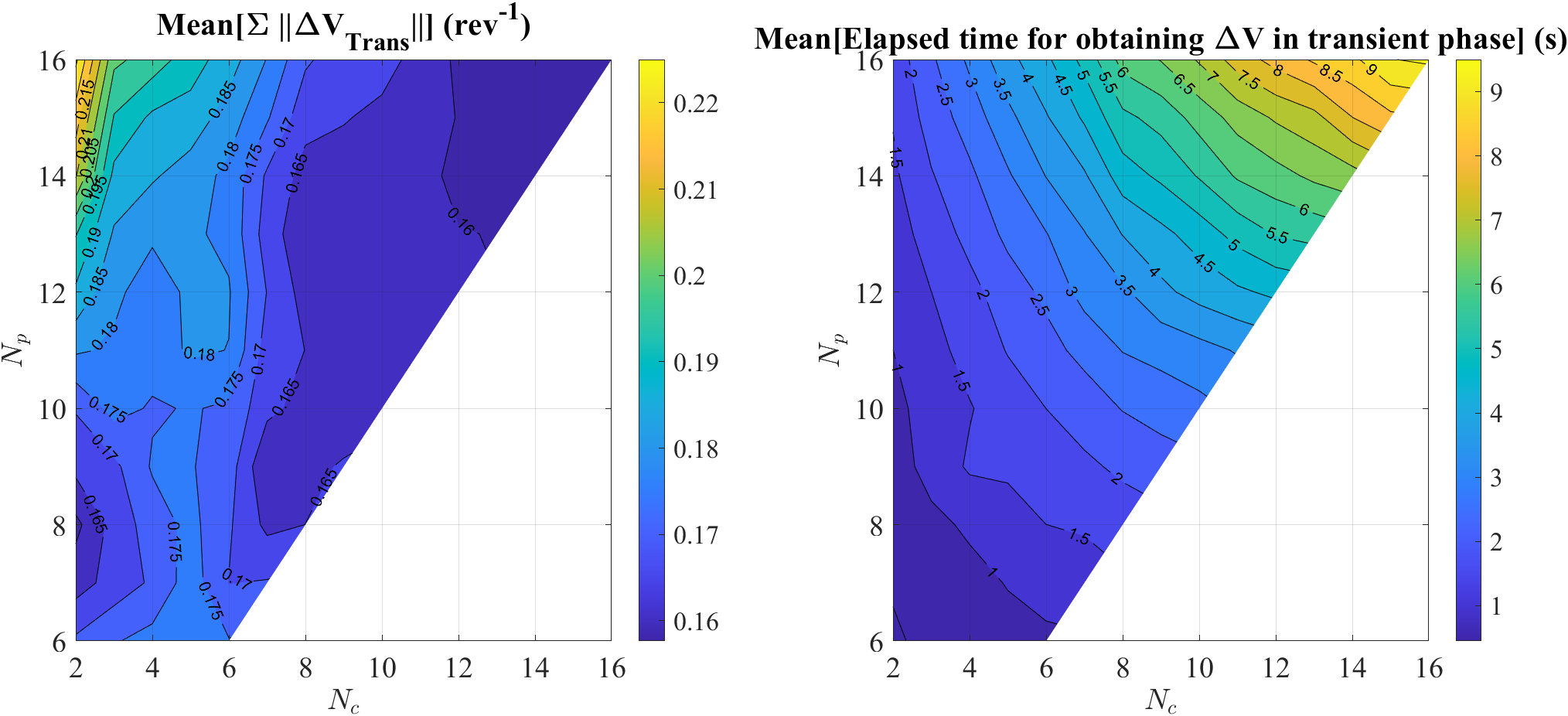}
    \caption{The control efforts and computation time for different \(N_p\) and \(N_c\) in case of NRHOs.}
    \label{fig:NRHO_contours}
\end{figure}

Figures~\ref{fig:LO_contours}–\ref{fig:NRHO_contours} show the average magnitude of the velocity impulse \(\Delta V\) and the average computation time per impulse for three different periodic orbit families. Each value is derived from a numerical simulation over 10 revolutions. The figures indicate that computational complexity scales approximately linearly with \(N_p\) and \(N_c\). However, the relationship between the control effort and the horizon lengths is not linear across all orbit families. {For LOs, the control effort peaks at $N_c = 4$, indicating that this is a non-optimal choice, while it reaches a minimum at $N_c = 2$. In contrast, for HOs, the control effort peaks at $N_c \in \{2, 3\}$ and $N_p \in [6, 8]$, suggesting that optimal performance is achieved with larger values of both $N_p$ and $N_c$. For NRHOs, the peak occurs at $N_p = 6$ and $N_c = 2$, implying that a larger gap between $N_p$ and $N_c$ leads to reduced system robustness.} This observation suggests that the NMPC scheme may lose robustness at large horizon values, as the associated optimization problem becomes significantly non-convex.


\section{Numerical Simulations}
The proposed NMPC system is evaluated through numerical simulations in a MATLAB environment, using the CasADi package to solve the online nonlinear optimization problem. The performance of this method is compared against two other approaches: the first is a standard NMPC scheme in which the reference trajectory is a predefined orbit, as embedded in common missions. The second is a similar scheme, where the parameter \(\chi\) varies at each time step along the prediction horizon. The proposed approach is tested on LOs, HOs, and NRHOs. To assess the robustness of the control system, the dynamic model includes an unexpected biased disturbance. {A biased disturbance is selected to simulate a strong perturbation, such as an unrealistic strong solar wind (for visualization and testing purposes), to prove the superiority of the proposed technique over the traditional one. The disturbance is assumed to be Gaussian with a zero mean and \(\sigma_q = 10^{-3}\). The initial state is assumed to be poorly estimated, with an initial state covariance of \(P = 10^{-6}\, I_6\).}

The prediction and control horizons for each periodic orbit family are selected based on Figures~\ref{fig:LO_contours}–\ref{fig:NRHO_contours}. The weighting matrices are manually tuned. The final values used in the simulations are: \(\mathbf{Q} = \mathbf{Q}_t = \text{diag}(1, 1, 1, 0, 0, 0)\), and \(\mathbf{R} = \text{diag}(10^{-2}, 10^{-2}, 10^{-2})\). In each orbit revolution, 20 control impulses are applied to the spacecraft.

\begin{figure}[H]
    \centering
    \includegraphics[width=\linewidth]{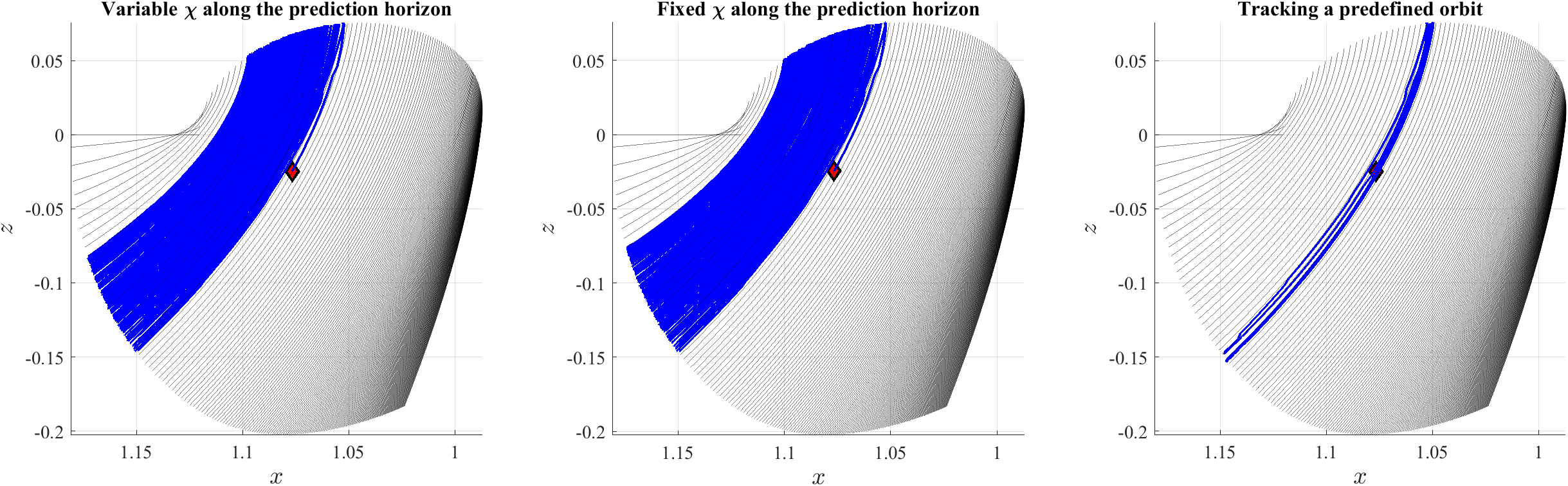}
    \caption{Spacecraft trajectory after 100 revolutions for the three control strategies in the case of HOs.}
    \label{fig:HO_compare}
\end{figure}

\begin{figure}[H]
    \centering
    \includegraphics[width=\linewidth]{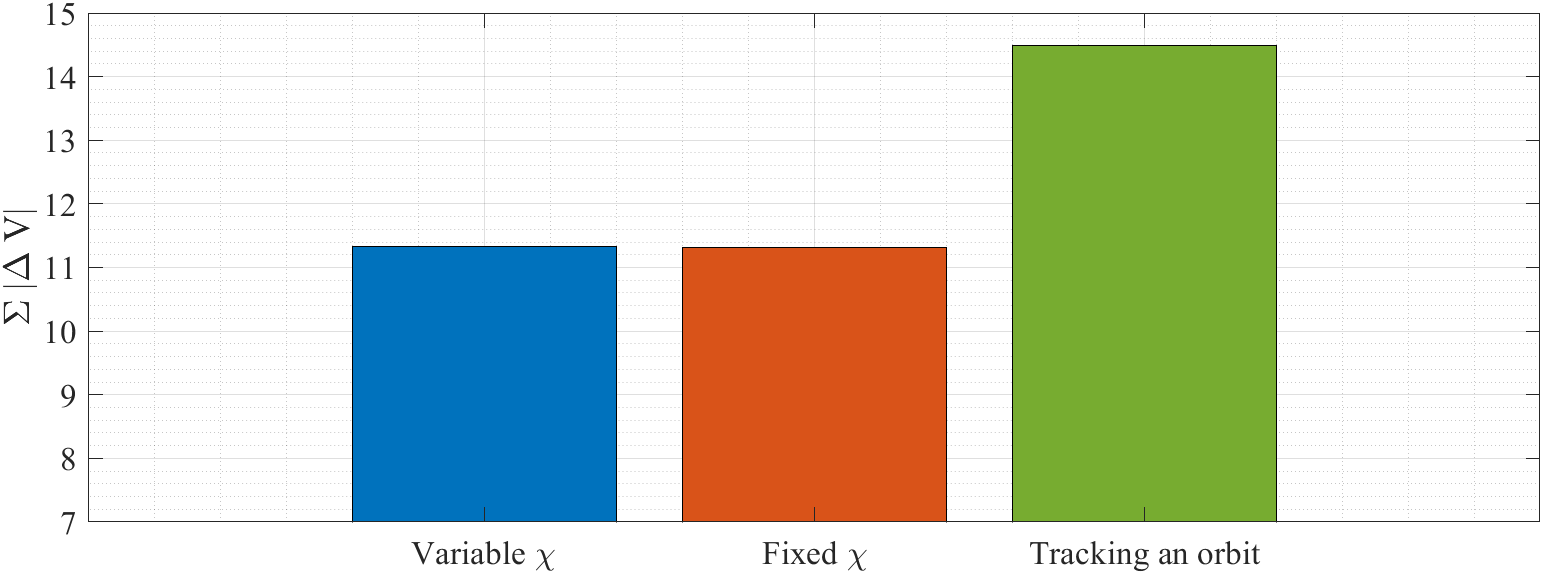}
    \caption{Total control effort after 100 revolutions for the three approaches in the case of HOs.}
    \label{fig:HO_controlEffort}
\end{figure}

Figure~\ref{fig:HO_compare} shows the spacecraft trajectory in the three test cases. In the first case, the parameter \(\chi\) varies at each time step along the prediction horizon. In the second case, \(\chi\) is treated as an optimization variable but is assumed constant along the horizon to simplify the problem. In the third case, the reference orbit is fixed, and the spacecraft is constrained to track only that orbit. Figure~\ref{fig:HO_controlEffort} illustrates the corresponding control efforts. The first two approaches yield similar control efforts, while the third requires significantly more. This suggests that the fixed-\(\chi\) approach is the most efficient, as it simplifies the optimization without increasing control demand.

\begin{figure}[H]
\centering
\includegraphics[width=\linewidth]{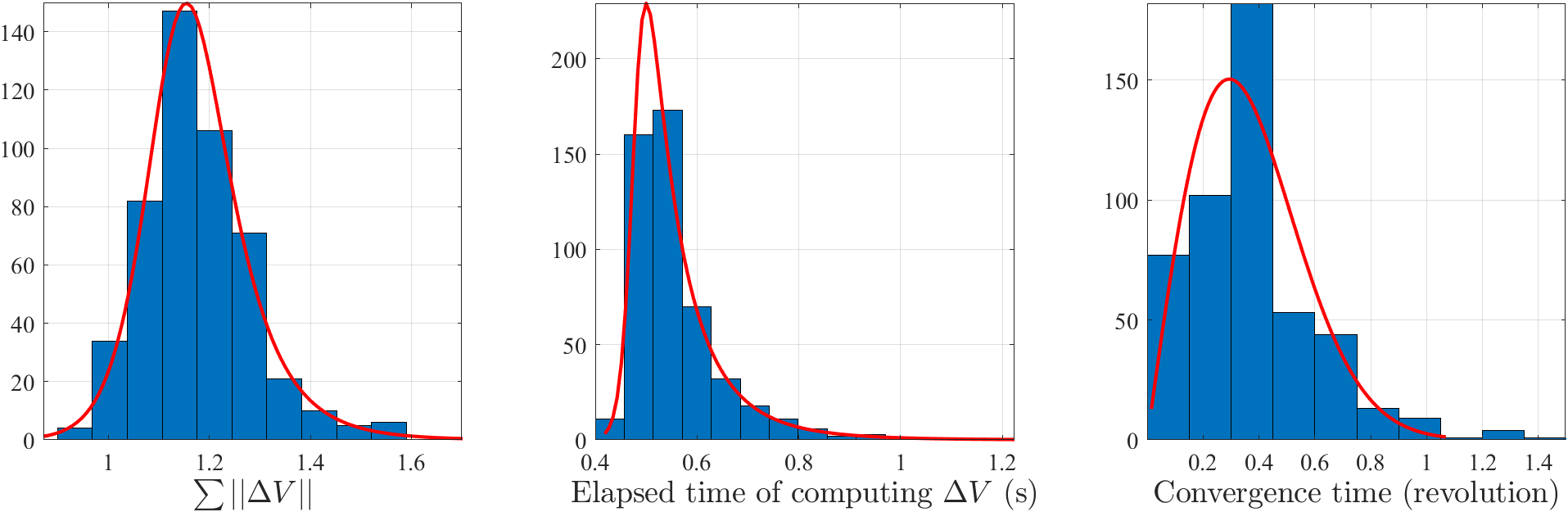}
\caption{Results of 500 Monte Carlo simulations for HOs.}
\label{fig:HO_MC}
\end{figure}

In order to verify the proposed NMPC approach, a sequence of 500 Monte Carlo simulations is conducted over 10 revolutions. In each simulation, the spacecraft starts from a different random initial state. The objectives are to ensure that the NMPC scheme can successfully track the orbit family, evaluate the convergence time to the steady state, assess the total control effort, and measure the computational time required to determine the optimal control input. Figure~\ref{fig:HO_MC} presents histograms of the results obtained from the Monte Carlo simulations. The results demonstrate successful tracking and convergence of the proposed NMPC scheme.

\begin{figure}[H]
    \centering
    \subfigure[LOs.]{%
        \includegraphics[width=0.45\textwidth]{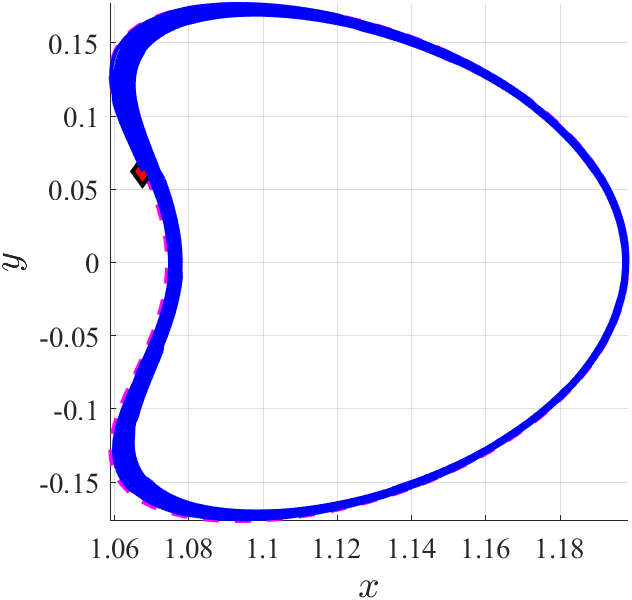}
        \label{fig:LO_compare}
    }
    \hfill
    \subfigure[NRHOs.]{%
        \includegraphics[width=0.45\textwidth]{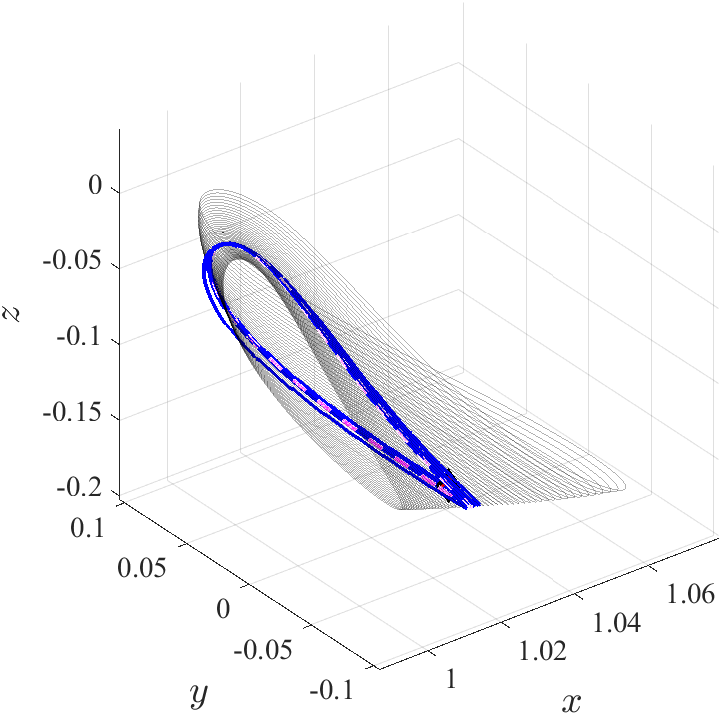}
        \label{fig:NRHO_compare}
    }
    \caption{Spacecraft trajectories for LOs and NRHOs under the proposed NMPC scheme.}
    \label{fig:LO_NRHO}
\end{figure}

To verify the applicability of the proposed method to other orbit families, it is also tested on LOs and NRHOs. The results, shown in Figure~\ref{fig:LO_NRHO}, indicate that the spacecraft remains within the orbit family in both cases, despite the presence of disturbances.

\begin{figure}[H]
    \centering
    \subfigure[Fixed-\(\chi\)]{%
        \includegraphics[width=0.45\textwidth]{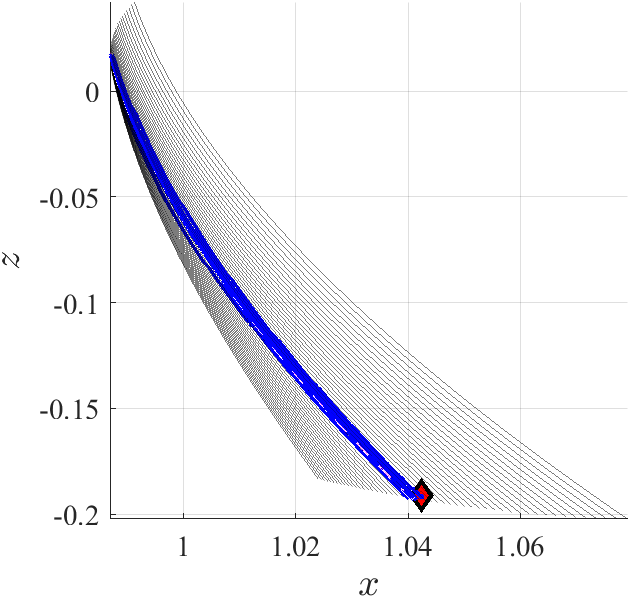}
        \label{fig:NRHO_traj_varChi}
    }
    \hfill
    \subfigure[Tracking an orbit]{%
        \includegraphics[width=0.45\textwidth]{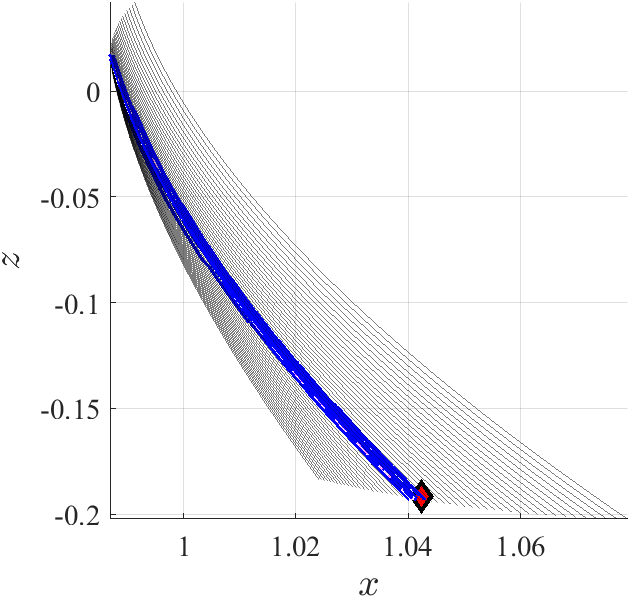}
        \label{fig:NRHO_traj_fixedChi}
    }
    \caption{Spacecraft trajectory after 20 revolutions for the tow control strategies
in the case of NRHOs.}
    \label{fig:NRHO_traj}
\end{figure}

\begin{figure}[H]
    \centering
    \includegraphics[width=0.6\linewidth]{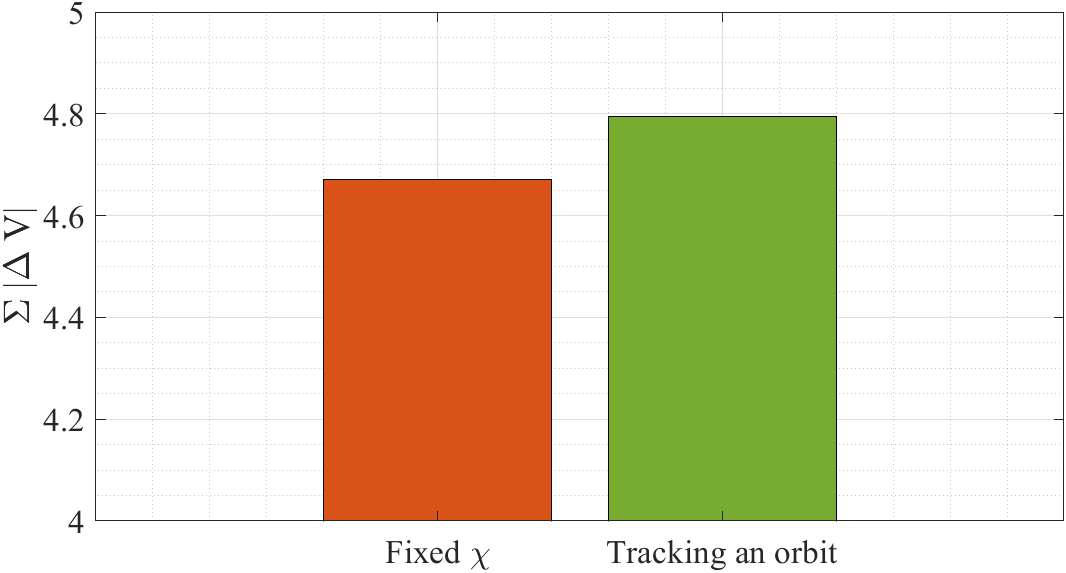}
    \caption{Total control effort after 20 revolutions for the three approaches in the case of NRHOs.}
    \label{fig:NRHO_controlEffort}
\end{figure}

{
Figure~\ref{fig:NRHO_traj} shows the spacecraft trajectory for two test cases, fixed-$\chi$ and predefined orbit tracking, assuming the system is subjected to a disturbance. Figure~\ref{fig:NRHO_controlEffort} illustrates the corresponding control efforts. This suggests that the fixed-$\chi$ approach is more efficient, similar to the HO case. While relative small over a period of 20 revolutions, the benefits of the approach are evident and they enable longer mission durations due to the reduced fuel consumption. 
}

\subsection{NMPC–EKF system}
\begin{figure}[H]
    \centering
    \includegraphics[width=0.7\linewidth]{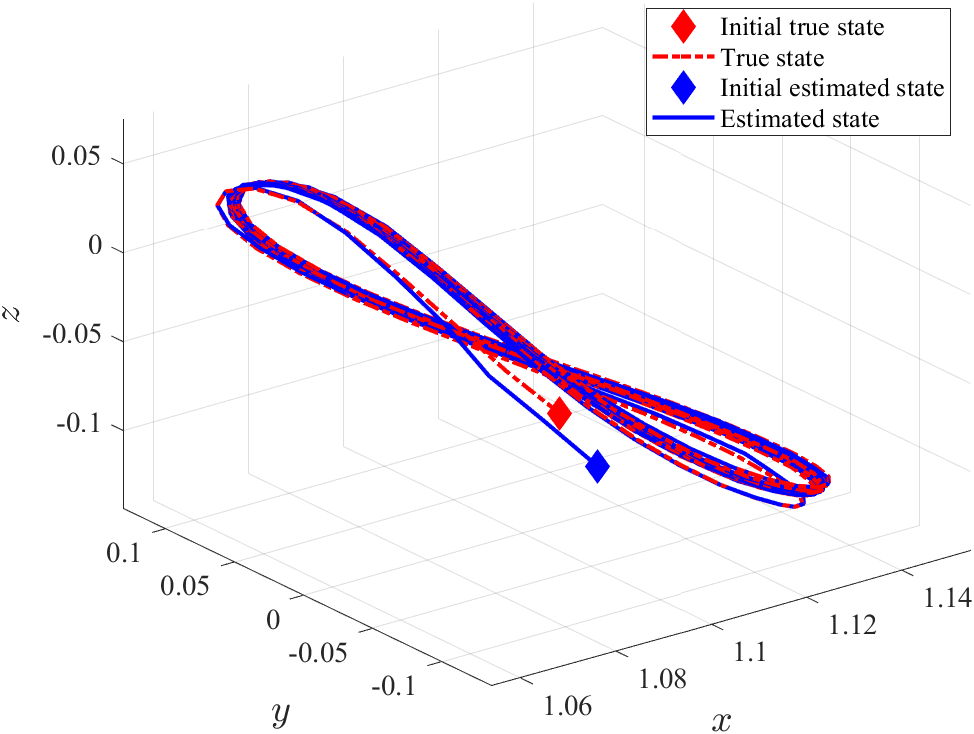}
    \caption{Spacecraft trajectory with integrated NMPC and EKF in the case of HOs.}
    \label{fig:HO_GNC}
\end{figure}

To further validate the proposed method, the NMPC controller is integrated with an observer. While several options, even optimized for the CRTBP [\citen{servadio2023uncertainty}], could be picked, a simple EKF has been chosen {to estimate the state due to its low computation efforts and being widely used in cislunar applications}, where the available measurements include the relative range and line-of-sight vector. {The EKF is responsible for estimating the states of the satellite at a given time step. These states are then used in the NMPC framework to compute the optimal reference orbit and the optimal velocity impulse to track that orbit. The numerical simulation verifies the EKF system by computing the error between the estimated and true states, while the NMPC system is verified by computing the error between the estimated and reference states.} Figure~\ref{fig:HO_GNC} compares the true and estimated spacecraft trajectories under the GNC system. The results demonstrate that the spacecraft remains within the periodic orbit family, confirming the effectiveness of the integrated NMPC–EKF approach.

\begin{figure}[H]
\centering
\includegraphics[width=\linewidth]{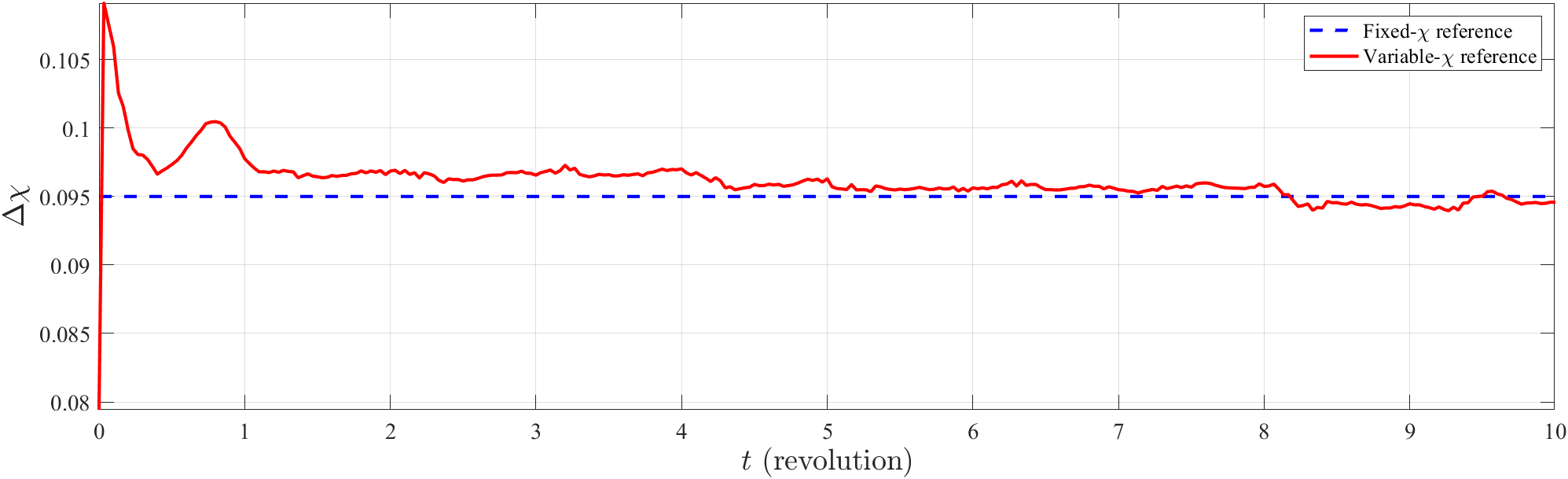}
\caption{Variation of $\Delta \chi$ over time for HOs.}
\label{fig:HO_chi}
\end{figure}

Figure~\ref{fig:HO_chi} shows the variation of $\Delta\chi$ over time. The parameter $\chi$ characterizes the spacecraft’s orbit. The figure indicates that the spacecraft requires approximately one revolution to reach a steady state. However, even in the steady state, the spacecraft does not remain in exactly the same orbit; instead, the orbit undergoes slight variations. This variation arises from tracking an orbit family rather than a single predefined orbit.

\begin{figure}[H]
\centering
\includegraphics[width=\linewidth]{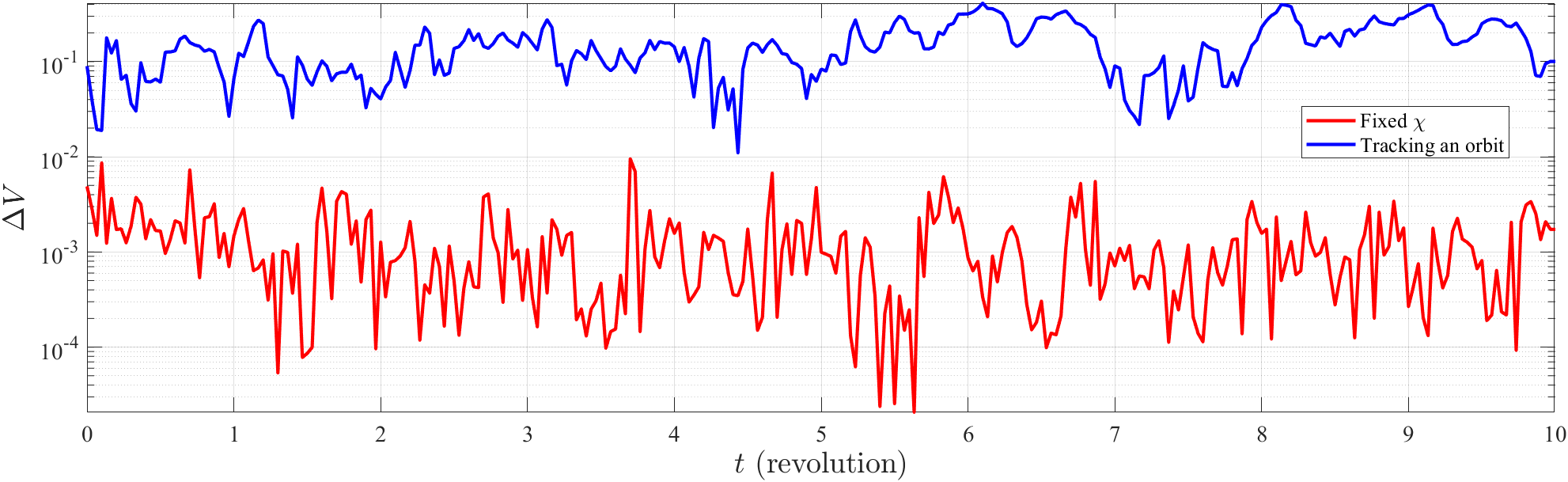}
\caption{Variation of $\Delta V$ over time for HOs.}
\label{fig:HO_dV}
\end{figure}

Figure~\ref{fig:HO_dV} presents the variation of $\Delta V$ over time. The control efforts do not converge to a steady-state value due to the presence of uncertainties introduced into the system dynamics. Compared to tactical tracking of a reference orbit, the newly developed approach requires significantly less control effort.

\begin{figure}[H]
\centering
\includegraphics[width=\linewidth]{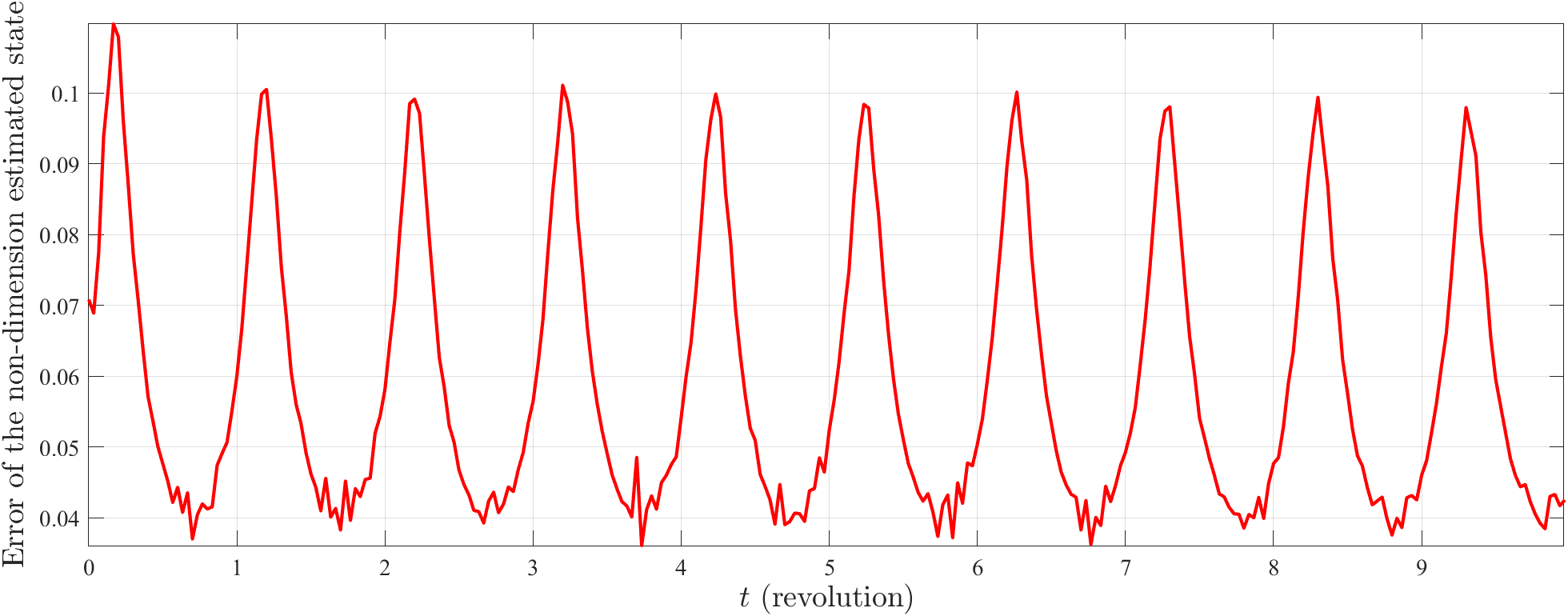}
\caption{Variation of the estimation error over time for HOs.}
\label{fig:HO_err}
\end{figure}

Figure~\ref{fig:HO_err} illustrates the variation of the error between the true and estimated states over time. The estimation error fluctuates around a certain mean value, indicating the stable performance of the EKF in this system.

\section{Conclusions}
This work introduced a novel NMPC framework for maintaining a spacecraft within a desired family of periodic orbits in the cislunar environment, without requiring strict adherence to a specific reference trajectory. By modeling spacecraft motion within the CR3BP and leveraging the PAC method, a comprehensive representation of orbit families was obtained. The parameterization of orbit members using two intuitive variables, combined with MPR, enabled efficient modeling of orbital states across the family. The developed NMPC strategy optimally selects and tracks trajectories within the family, while minimizing fuel usage through impulsive control actions. The integration of an EKF observer further enhanced the robustness of the control system by providing accurate relative state estimates. Numerical simulations across Lyapunov, halo, and near-rectilinear halo orbits near \(\mathcal{L}_1\) and \(\mathcal{L}_2\) validated the effectiveness of the proposed method. The results revealed substantial fuel savings and trajectory flexibility compared to conventional tracking approaches, highlighting the potential of this approach for future cislunar mission design and stationkeeping applications. Future work may include advanced filtering technique, such as those proposed in~[\citen{servadio2020nonlinear, servadio2020recursive, servadio2021differential, servadio2022maximum, servadio2022nonlinear, servadio2024likelihood}], to further enhance the accuracy of the estimated trajectory, particularly in scenarios where the initial state is poorly known.


\bibliographystyle{ieeetr}
\bibliography{references}

\begin{thebibliography}{10}

\bibitem{koon2000dynamical}
W.~S. Koon, M.~W. Lo, J.~E. Marsden, and S.~D. Ross, ``Dynamical systems, the three-body problem and space mission design,'' in {\em Equadiff 99: (In 2 Volumes)}, pp.~1167--1181, World Scientific, 2000.

\bibitem{wilmer2024preliminary}
A.~P. Wilmer, R.~A. Bettinger, L.~M. Shockley, and M.~J. Holzinger, ``Preliminary investigation and proposal of periodic orbits and their utilization for logistics in the cislunar regime,'' {\em Space Policy}, p.~101635, 2024.
\newblock https://doi.org/10.1016/j.spacepol.2024.10163.

\bibitem{woodard2009artemis}
M.~Woodard, D.~Folta, and D.~Woodfork, ``Artemis: the first mission to the lunar libration orbits,'' in {\em 21st International Symposium on Space Flight Dynamics, Toulouse, France}, 2009.

\bibitem{merri2018lunar}
M.~Merri and M.~Sarkarati, ``Lunar orbiter platform-gateway: a clear use case for ccsds mo services,'' in {\em 2018 AIAA SPACE and Astronautics Forum and Exposition}, p.~5337, 2018.
\newblock https://doi.org/10.2514/6.2018-5337.

\bibitem{national2016nasa}
N.~Aeronautics, {\em NASA's journey to Mars: pioneering next steps in space exploration}.
\newblock Government Printing Office, 2016.

\bibitem{baker2024comprehensive}
B.~Baker-McEvilly, S.~Bhadauria, D.~Canales, and C.~Frueh, ``A comprehensive review on cislunar expansion and space domain awareness,'' {\em Progress in Aerospace Sciences}, vol.~147, p.~101019, 2024.
\newblock https://doi.org/10.1016/j.paerosci.2024.10101.

\bibitem{whitley2016options}
R.~Whitley and R.~Martinez, ``Options for staging orbits in cislunar space,'' in {\em 2016 IEEE Aerospace Conference}, pp.~1--9, IEEE, 2016.
\newblock https://doi.org/10.1109/AERO.2016.7500635.

\bibitem{richardson1980analytic}
D.~L. Richardson, ``Analytic construction of periodic orbits about the collinear points,'' {\em Celestial mechanics}, vol.~22, no.~3, pp.~241--253, 1980.
\newblock https://doi.org/10.1007/BF01229511.

\bibitem{pritchett2018impulsive}
R.~E. Pritchett, E.~Zimovan, and K.~Howell, ``Impulsive and low-thrust transfer design between stable and nearly-stable periodic orbits in the restricted problem,'' in {\em 2018 Space Flight Mechanics Meeting}, p.~1690, 2018.
\newblock https://doi.org/10.2514/6.2018-1690.

\bibitem{singh2020low}
S.~K. Singh, B.~D. Anderson, E.~Taheri, and J.~L. Junkins, ``Low-thrust earth-moon transfers via manifolds of a halo orbit in the cis-lunar space,'' in {\em 43rd Annual AAS Guidance, Navigation and Control Conference}, 2020.

\bibitem{davis2017stationkeeping}
D.~C. Davis, S.~M. Phillips, K.~C. Howell, S.~Vutukuri, and B.~P. McCarthy, ``Stationkeeping and transfer trajectory design for spacecraft in cislunar space,'' in {\em AAS/AIAA Astrodynamics Specialist Conference}, vol.~8, Springer Nature London, 2017.

\bibitem{van2016tadpole}
A.~G. Van~Anderlecht, ``Tadpole orbits in the l4/l5 region: Construction and links to other families of periodic orbits,'' Master's thesis, Purdue University, 2016.

\bibitem{wilmer2021cislunar}
A.~Wilmer, R.~A. Bettinger, and B.~Little, {\em Cislunar Periodic Orbit Constellation Assessment for Space Domain Awareness of L1 and L2 Halo Orbits}, p.~4191.
\newblock 2021.
\newblock https://doi.org/10.2514/6.2021-4191.

\bibitem{fay2024investigation}
T.~J. Fay, A.~P. Wilmer, and R.~A. Bettinger, ``Investigation of near-rectilinear halo orbit search and rescue using staging l1/l2 lyapunov and distant retrograde orbit families,'' {\em Journal of Space Safety Engineering}, 2024.
\newblock https://doi.org/10.1016/j.jsse.2024.04.009.

\bibitem{negri2020generalizing}
R.~B. Negri and A.~F. Prado, ``Generalizing the bicircular restricted four-body problem,'' {\em Journal of Guidance, Control, and Dynamics}, vol.~43, no.~6, pp.~1173--1179, 2020.
\newblock https://doi.org/10.2514/1.G004848.

\bibitem{oshima2022multiple}
K.~Oshima, ``Multiple families of synodic resonant periodic orbits in the bicircular restricted four--body problem,'' {\em Advances in Space Research}, vol.~70, no.~5, pp.~1325--1335, 2022.
\newblock https://doi.org/10.1016/j.asr.2022.06.009.

\bibitem{wilmer2021lagrangian}
A.~P. Wilmer and R.~A. Bettinger, ``Lagrangian derivation and stability analysis of multi-body gravitational dynamical models with application to cislunar periodic orbit propagation,'' in {\em Astrodyn. Spec. Conf}, 2021.

\bibitem{salzo2023design}
F.~P. Salzo, G.~Bucchioni, R.~Vazquez, {\em et~al.}, ``Design of a model predictive controller for formation flight on quasi-halo orbits,'' 2023.

\bibitem{sanchez2020chance}
J.~C. Sanchez, F.~Gavilan, and R.~Vazquez, ``Chance-constrained model predictive control for near rectilinear halo orbit spacecraft rendezvous,'' {\em Aerospace Science and Technology}, vol.~100, p.~105827, 2020.
\newblock https://doi.org/10.1016/j.ast.2020.105827.

\bibitem{capannolo2023model}
A.~Capannolo, G.~Zanotti, M.~Lavagna, and G.~Cataldo, ``Model predictive control for formation reconfiguration exploiting quasi-periodic tori in the cislunar environment,'' {\em Nonlinear Dynamics}, vol.~111, no.~8, pp.~6941--6959, 2023.
\newblock https://doi.org/10.1007/s11071-022-08214-8.

\bibitem{quartullo2023periodic}
R.~Quartullo, A.~Garulli, and I.~Kolmanovsky, ``Periodic model predictive control for tracking halo orbits in the elliptic restricted three-body problem,'' {\em IEEE Transactions on Control Systems Technology}, vol.~31, no.~5, pp.~1971--1981, 2023.
\newblock https://doi.org/10.1109/TCST.2023.3291548.

\bibitem{berning2020suboptimal}
A.~W. Berning~Jr, D.~Liao-McPherson, A.~Girard, and I.~Kolmanovsky, ``Suboptimal nonlinear model predictive control strategies for tracking near rectilinear halo orbits,'' {\em arXiv preprint arXiv:2008.09240}, 2020.
\newblock https://doi.org/10.48550/arXiv.2008.09240.

\bibitem{servadio2021koopman}
S.~Servadio, D.~Arnas, and R.~Linares, ``A koopman operator tutorial with othogonal polynomials,'' {\em arXiv preprint arXiv:2111.07485}, 2021.
\newblock https://doi.org/10.48550/arXiv.2111.07485.

\bibitem{servadio2022dynamics}
S.~Servadio, D.~Arnas, and R.~Linares, ``Dynamics near the three-body libration points via koopman operator theory,'' {\em Journal of Guidance, Control, and Dynamics}, vol.~45, no.~10, pp.~1800--1814, 2022.
\newblock https://doi.org/10.2514/1.G006519.

\bibitem{henon1969numerical}
M.~H{\'e}non, ``Numerical exploration of the restricted problem, v,'' {\em Astronomy and Astrophysics, vol. 1, p. 223-238 (1969).}, vol.~1, pp.~223--238, 1969.

\bibitem{breakwell1979halo}
J.~V. Breakwell and J.~V. Brown, ``The ‘halo’family of 3-dimensional periodic orbits in the earth-moon restricted 3-body problem,'' {\em Celestial mechanics}, vol.~20, no.~4, pp.~389--404, 1979.
\newblock https://doi.org/10.1007/BF01230405.

\bibitem{atallah2024advances}
M.~Atallah and S.~Servadio, ``Advances in cislunar periodic solutions via taylor polynomial maps,'' {\em arXiv preprint arXiv:2409.03692}, 2024.
\newblock https://doi.org/10.48550/arXiv.2409.03692.

\bibitem{andersson2019casadi}
J.~A. Andersson, J.~Gillis, G.~Horn, J.~B. Rawlings, and M.~Diehl, ``Casadi: a software framework for nonlinear optimization and optimal control,'' {\em Mathematical Programming Computation}, vol.~11, pp.~1--36, 2019.
\newblock https://doi.org/10.1007/s12532-018-0139-4.

\bibitem{elhesasy2023non}
M.~Elhesasy, T.~N. Dief, M.~Atallah, M.~Okasha, M.~M. Kamra, S.~Yoshida, and M.~A. Rushdi, ``Non-linear model predictive control using casadi package for trajectory tracking of quadrotor,'' {\em Energies}, vol.~16, no.~5, p.~2143, 2023.
\newblock https://doi.org/10.3390/en16052143.

\bibitem{atallah2024analytic}
M.~Atallah, M.~Okasha, and O.~Abdelkhalik, ``Analytic optimal control for multi-satellite assembly using linearized twistor-based model,'' {\em Advances in Space Research}, vol.~74, no.~10, pp.~5142--5155, 2024.
\newblock https://doi.org/10.1016/j.asr.2024.08.072.

\bibitem{atallah2023development}
M.~A. Atallah, ``Development of guidance, navigation, and control systems for multi-spacecraft assembly in proximity operations,'' 2023.

\bibitem{servadio2023uncertainty}
S.~Servadio, W.~Parker, and R.~Linares, ``Uncertainty propagation and filtering via the koopman operator in astrodynamics,'' {\em Journal of Spacecraft and Rockets}, vol.~60, no.~5, pp.~1639--1655, 2023.
\newblock https://doi.org/10.2514/1.A35688.

\bibitem{servadio2020nonlinear}
S.~Servadio, R.~Zanetti, and B.~A. Jones, ``Nonlinear filtering with a polynomial series of gaussian random variables,'' {\em IEEE Transactions on Aerospace and Electronic Systems}, vol.~57, no.~1, pp.~647--658, 2020.
\newblock https://doi.org/10.1109/TAES.2020.3028487.

\bibitem{servadio2020recursive}
S.~Servadio and R.~Zanetti, ``Recursive polynomial minimum mean-square error estimation with applications to orbit determination,'' {\em Journal of Guidance, Control, and Dynamics}, vol.~43, no.~5, pp.~939--954, 2020.
\newblock https://doi.org/10.2514/1.G004544.

\bibitem{servadio2021differential}
S.~Servadio and R.~Zanetti, ``Differential algebra-based multiple gaussian particle filter for orbit determination,'' {\em Journal of Optimization Theory and Applications}, vol.~191, no.~2, pp.~459--485, 2021.
\newblock https://doi.org/10.1007/s10957-021-01934-8.

\bibitem{servadio2022maximum}
S.~Servadio, R.~Zanetti, and R.~Armellin, ``Maximum a posteriori estimation of hamiltonian systems with high order taylor polynomials,'' {\em The Journal of the Astronautical Sciences}, vol.~69, no.~2, pp.~511--536, 2022.
\newblock https://doi.org/10.1007/s40295-022-00304-4.

\bibitem{servadio2022nonlinear}
S.~Servadio, F.~Cavenago, P.~Di~Lizia, and M.~Massari, ``Nonlinear prediction in marker-based spacecraft pose estimation with polynomial transition maps,'' {\em Journal of Spacecraft and Rockets}, vol.~59, no.~2, pp.~511--523, 2022.
\newblock https://doi.org/10.2514/1.A35068.

\bibitem{servadio2024likelihood}
S.~Servadio, ``Likelihood scouting via map inversion for a posterior-sampled particle filter,'' {\em IEEE Transactions on Aerospace and Electronic Systems}, 2024.
\newblock https://doi.org/10.1109/TAES.2024.3446757.

\end{thebibliography}

\end{document}